%% file: wikisoc.tex
\documentclass[preprints,article,accept,moreauthors,pdftex]{Definitions/mdpi} 

\pdfoutput=1

\usepackage{amssymb}
\usepackage{amsmath}

\usepackage{color}
\usepackage{soul}

\newcommand\GR{G_{\rm R}}
\newcommand\Grr{{G_{\rm rr}}}
\newcommand\Gss{G_{\rm ss}}
\newcommand\Grs{{G_{\rm rs}}}
\newcommand\Gsr{{G_{\rm sr}}}
\newcommand\Gqr{{G_{\rm qr}}}
\newcommand\Gpr{{G_{\rm pr}}}

\newcommand\Gqrnd{{G_{\rm qr}^{\rm (nd)}}}
\newcommand\Wqr{{W_{\rm qr}}}
\newcommand\Wpr{{W_{\rm pr}}}
\newcommand\Wrr{{W_{\rm rr}}}
\newcommand\Wqrnd{{W_{\rm qr}^{\rm (nd)}}}
\newcommand\WR{{W_{\rm R}}}
\newcommand{\Nr}{N_{\rm r}}
\newcommand{\Ns}{N_{\rm s}}
\newcommand{\PR}{P_{\rm r}}
\newcommand{\PS}{P_{\rm s}}

\firstpage{1} 
\pubvolume{xx}
\issuenum{1}
\articlenumber{5}
\pubyear{2024}
\copyrightyear{2024}
\history{Received: December 4, 2024; Accepted: date; Published: date}

\Title{Relations of society concepts and religions\\
        from Wikipedia networks}

\Author{Klaus M. Frahm
$^{1}$\orcidA{},   and 
Dima~L.~Shepelyansky$^{1,*}$\orcidB{}}

\AuthorNames{K.M.Frahm and D.L.Shepelyansky}

\address{%
$^{1}$ \quad Laboratoire de Physique Th\'eorique,  
Universit\'e de Toulouse, CNRS, UPS, 31062 Toulouse, France\\
}

\corres{Correspondence: dima@irsamc.ups-tlse.fr; Tel. +33-56155-60-68 (D.L.S.)}

\abstract{We analyze the Google matrix 
of directed networks 
of Wikipedia articles related to 
8 recent Wikipedia language editions representing 
different cultures 
(English, Arabic, German, Spanish,
French, Italian, Russian, Chinese). 
Using the reduced Google matrix algorithm
we determine relations and interactions
of 23 society concepts and
17 religions represented by
their respective articles for each of the 8 editions.
The effective Markov transitions are found to be more 
intense inside the two blocks of society concepts
and religions while transitions between
the blocks are significantly reduced.
We establish 5 poles of influence for society
concepts (Law, Society, Communism,
Liberalism, Capitalism)
as well as 5 poles for religions
(Christianity, Islam, Buddhism,
Hinduism, Chinese folk religion)
and determine how they affect other entries. 
We compute inter edition correlations for different 
key quantities providing a quantitative analysis of 
the differences or the proximity
of views of the 8 cultures with respect to the 
selected society concepts and religions.
}
  

\begin{document}
\section{Introduction}
\label{sec1}

During human society evolution several concepts of society
have been developed. Their relations and interactions
are studied by social sciences as discussed for example 
in \cite{jarvie,gellner}.
Among these concepts there are those linked to social society structure,
as e.g. law, education, culture;
political formations as e.g. capitalism, socialism, communism and others.
These concepts represent social concepts of society.
Other  concepts are represented by religions which play
an important role in human society formation.
This role and relations of religions and human
societies have been studied and investigated from various view points 
including history, philosophy, 
psychology, archaeology, economy and other sciences (see e.g. 
\cite{casanova,reese,barrett,whitehouse,atran,boyer,hopfe} and Refs. therein).
However, the mathematical analysis of these relations is essentially absent 
since it is rather difficult to use and apply rigorous mathematical 
methods to human sciences and religions. In this work, we try to develop
a mathematical framework to investigate  religions and society 
concepts relations based on the Wikipedia network analysis. 
We consider a directed network composed of
Wikipedia articles (nodes) with generated hyperlinks (citations) between them.
Already two decades ago the free online Wikipedia  has superseded old 
encyclopedias such as Encyclopedia Britannica \cite{britanica} 
in quality and volume of articles related to scientific topics 
\cite{wiki1}. At present the academic analysis of information contained 
in Wikipedia finds more and more applications as reviewed and
described in \cite{reagle,finn,abramowicz,wikiacad3,wikiacad4}.
Thus Wikipedia contains an enormous diverse database of human knowledge
which we use to determine relations of religions and society concepts
obtained on purely mathematical grounds. Of course, religions are also human
concepts related to society but for clarity
we consider two groups of concepts formed by society: society
social concepts and religions.

We analyze Wikipedia networks of different language editions
by the Google matrix approach and Page Rank algorithm, originally developed 
and applied by Brain and Page
for the World Wide Web search analysis \cite{brin,meyer}.
We also use additional algorithms such as
CheiRank  \cite{cheirank,wikizzs,rmp2015} and the reduced Google matrix (REGOMAX)
\cite{politwiki}. Various applications of these algorithms with an analysis 
of the importance and interactions of various Wikipedia articles and groups 
of articles have been reported for historical figures \cite{wikizzs,eomwiki24},
world universities \cite{wikizzs,wrwu2017}, politicians \cite{politwiki},
banks and world countries  \cite{wikibank} and other examples. 
The REGOMAX algorithm is particularly useful since it
allows to determine effective interactions between nodes of a selected group 
taking into account direct and all indirect pathways between these nodes 
via the global network with a size exceeding by orders of magnitude
the number of selected group nodes. These tools have been shown to operate 
efficiently not only for Wikipedia networks but also for the world trade 
network of UN COMTRADE  \cite{wtn3} and the MetaCore network of 
protein-protein interactions \cite{grmetacore}.

In this work, we focus on the directed networks 
obtained from 8 actual Wikipedia language editions:
English (EN), Arab (AR), German (DE), Spanish (ES), French (FR),
Italian (IT), Russian (RU), Chinese (ZH).
We consider that these 8 networks describe
8 different cultures since a language is
one of the most important elements of culture.
The 8 networks are generated from Wikipedia edition dumps of 1 October 2024
and the main network parameters are given in Table~\ref{table1}.

\begin{table}
\caption{\label{table1} Table of the 8 used Wikipedia editions (code 
in 2nd column) in different languages (1st column) and corresponding 
network data where $N$ denotes the number of network nodes (3rd column), 
$N_\ell$ the total number of links (4th column) and $N_d$ the 
number of dangling nodes (5th column). The last column provides 
the ratio $N_\ell/N$. 
All networks were created from Wikipedia xml-dump files dated 
from October 1, 2024 (excluding redirection and special technical 
nodes and links). 
}
\centering
\include{table1}
\end{table}

For our analysis we select from Wikipedia 40 articles about human concepts 
composed of 23 society concepts and 17 religions (and branches 
of religions) given in Table~\ref{table2}.
These 40 concepts are analyzed for each of 8 Wikipedia edition networks.
We show that the REGOMAX algorithm allows to determine 
nontrivial interactions among the 23 selected society concepts,
17 religions and also relations between these two groups
for each of the 8 networks of Table~\ref{table1}.
Furthermore, the obtained results also allow to characterize
quantitatively the correlations between these 8 cultures.
 
The paper is composed as follows: in Section~\ref{sec2}, we 
present some technical details about the used data sets 
and in Section~\ref{sec3} 
the used mathematical tools for network and correlation analysis 
are briefly reviewed. 
Section~\ref{sec4} presents and provides a detailed discussion of our 
results and Section~\ref{sec5} provides the final discussion and conclusion.

Additional materiel, additional figures and data files are 
available at \cite{ourwebpage}.

\begin{table}
\caption{\label{table2} Table of the subset of $N_r=40$ 
selected Wikipedia articles (nodes) from the English Wikipedia 
edition EN with 23 society articles 
($K_g=1,\ldots,23$) and 17 religion articles ($K_g=24,\ldots,40$)
both separated by an additional horizontal line. 
All data in this table refer to the English Wikipedia edition EN. 
Here $K_g$ represents the global index of this group (1st column) 
obtained by first PageRank ordering the 23 society nodes 
and subsequently PageRank ordering the 17 religion nodes. 
The other columns correspond 
to the group-local $K$- and $K^*$-rank indices (2nd and 3rd columns), 
the exact (English) Wikipedia article title (4th column), 
a short two character code for each article (5th column), 
network global $K_{\rm M}$- and $K_{\rm M}^*$-rank indices 
(6th and 7th columns) 
and a subgroup (pole) index (8th column). 
In this table PageRank ordering, $K$- and $K^*$-indices were computed 
using the English Wikipedia edition EN. 
The two character code for each article is by default obtained from the 
first two characters of the title with some modifications for the 2nd 
character to avoid double codes or one of the 8 codes already used for the 
8 Wikipedia language editions (see 2nd column in Table~\ref{table1}). 
The subgroup index defines for each society and religion block 
five subgroups and the label ``(T)'' defines a top pole node for each subgroup. 
The two character codes, the pole index and top node label are used 
later in the network diagrams presented and discussed  
in Subsection \ref{subsec4.3}.
}
\centering
\include{table2}
\end{table}

\section{Data sets of Wikipedia networks}
\label{sec2}

\subsection{Construction of the Wikipedia networks}

To construct the Wikipedia networks for the different language editions 
of Table~\ref{table1}, 
we downloaded the official xml-dump files for each edition 
(versions with all pages and articles in a single file) with 
time stamp of October 1, 2024. First, we extracted for a given edition 
the titles of all main content articles with namespace key value being 0 
thus excluding technical Wikipedia articles of type Category, Help, 
File, Wikemedia etc. having other namespace key values. 
Also titles corresponding to a redirection to another 
existing article were not taken into account. This procedure provides the 
list of all article titles/network nodes 
with a technical index given by the initial order 
of the xml-dump file which is typically roughly alphabetical (but not 
exactly). This technical initial index, while being important 
to store the network links and for the different computation codes, 
will not be used 
in this work to characterize specific articles/nodes. Instead, 
we will use the PageRank index $K_M$ (see below) to characterize 
different nodes. 

In a second step, we determined for each article the links to other 
articles in the same Wikipedia edition. These links are (with some 
technical complications) available in the xml-dump files as clear text 
article titles but sometimes it is necessary to capitalize the first 
character of a link name to identify the proper article title, which 
is always capitalized for the first character, and the 
corresponding node index. This capitalization procedure is for accented 
or special characters (with multibyte utf8 codes), e.g. in the 
case of French or Russian article titles, a technical nontrivial operation 
for which we used the library Utf8proc \cite{utf8proc}.
Links to redirection titles, outside webpages or other wikipedia editions 
were not taken into account. 
Multiple links between two articles, in particular several 
links pointing to a particular 
position inside the target article, were counted as a single unique link.
Links from an article to itself were kept but also only as a single link.

In this way, we obtained the Wikipedia networks with full ``official'' 
lists of article titles for the 8 editions and tables of links, 
with some key values 
shown in Table~\ref{table1}, e.g. with typical node 
numbers between $N\approx 6.9\times 10^6$ (for the English Wikipedia edition 
EN) and $N\approx 1.2\times 10^6$ (for the Arabic Wikipedia edition AR) and 
typical link numbers between 
$N_\ell\approx 1.9\times 10^8$ (EN) 
and $N\approx 1.6\times 10^7$ (AR). 
The ratio $N_\ell/N$ between both has typical values $13$-$27$.
Table~\ref{table1} also shows the number of dangling nodes $N_d$ corresponding 
to nodes having no outgoing link to another Wikipedia article of the 
same edition. These nodes require special treatment in the Google matrix 
approach (see next section) and the typical values are $N_d\sim 10^3$-$10^5$ 
which represents a very small fraction of the full number of nodes 
$N\sim 10^6$-$10^7$. 

\subsection{Edition specific subsets}
\label{subsec2.2}

Starting with English Wikipedia, we select a group of 40 articles 
about 23 society concepts and 17 religions (or branches of religions) 
shown in Table~\ref{table2}. The order of these articles is fixed 
by first PageRank ordering the 23 society nodes and then the 17 religion 
nodes providing the group index $K_g$. 
Then for each of the other editions, we determine an equivalent group using 
the official Wikipedia links between two different editions. 
For this we downloaded the 
source html files of the 40 EN-articles and extracted from these files 
the corresponding article titles of the other editions. 
This information is indeed contained in these source files in the 
other language submenu where the corresponding articles in other language 
editions can be selected from a given English Wikipedia page. It turns 
out that for all the other 7 language editions and each of 
the 40 EN-articles of Table~\ref{table2}, we were able to find the 
corresponding article associated to the other edition (for some 
other language Wikipedia editions, not used in this work here, the number 
of found ``translated'' articles may be less than 40). 

We note that this procedure is important and that simple linguistic 
title translation may lead to wrong articles in the other edition. 
For example the EN-article ``Society'' ($K_g=15$ in Table~\ref{table2}) 
translates linguistically 
to ``Gesellschaft'' in German but the DE-article ``Gesellschaft'' 
corresponds more to a technical article with links to several other 
DE-articles using the word ``Gesellschaft'' in the title and for different 
contexts (e.g. sociology, ethonology, state-law etc.). However, 
the official inter edition Wikipedia 
link from the EN-article ``Society'' to the DE-Wikipedia edition 
provides the title ``Gesellschaft (Soziologie)'' which is a different 
DE-Wikipedia article than ``Gesellschaft''. There are some other similar 
examples like this, also for FR-Wikipedia. In particular both articles 
about the Orthodox Church with very specific differences between them 
(with $K_g=29,\,40$ in Table~\ref{table2}), 
require to use the official inter edition Wikipedia links while 
simple linguistic translation may easily lead to the wrong article. 

In this work, we will for convenience always use (in Tables and 
Figures below) the English article titles given in Table~\ref{table2}, even 
when we speak about another edition. Of course, the network 
index values of the 40 group nodes, necessary for the computation of 
the reduced Google matrix (see below), were correctly determined 
individually for each edition using the properly translated group list 
for the same edition, eventually using article titles in special 
character sets (especially for AR, RU, ZH). 

\begin{table}
\caption{\label{table3} Table of local $K$- and $K^*$-indices 
of the subset of $N_r=40$ selected Wikipedia articles (nodes) 
of Table~\ref{table2} obtained from the networks of all 8 Wikipedia 
editions of Table~\ref{table1}. For each edition 
(other than EN) a subset of $N_r=40$ articles was selected by 
using the official Wikipedia translation of the titles 
of Table~\ref{table2} for EN to the titles of the corresponding edition 
in the other language (AR to ZH). For each edition specific group the reduced 
matrix $G_R$, PageRank and CheiRank vector were computed using the 
corresponding edition Wikipedia network providing 
local $K$- and $K^*$-indices visible with two values $K;K^*$ per 
entry in columns 3 to 10 (for the 8 editions). 
The columns 1 and 2 provide the index $K_g$ and the short 
code for each node defined in Table~\ref{table2}. 
The additional horizontal line separates the society nodes ($K_g\le 23$) 
from the religion nodes ($K_g\ge 24$). 
The group nodes of each edition are 
also visible in Figs. \ref{fig1} and \ref{fig2} showing the global 
Wikipedia network structure for each edition in the 
$\ln(K_{\rm M})$-$\ln(K_{\rm M}^*)$ plane.
}
\centering
\include{table3}
\end{table}

In Table~\ref{table3}, we summarize the group local PageRank and CheiRank 
indices $K$ and $K^*$ (see next section) for 
all 40 subset nodes and all 8 Wikipedia editions of Table~\ref{table1} 
and using for each edition the properly translated group for this edition 
as described above.

\section{Google matrix algorithms} 
\label{sec3}

\subsection{Google matrix construction}
We construct the Google matrix $G$ of the different 
Wikipedia networks with $N$ nodes (articles) in the usual way 
\cite{brin,meyer,rmp2015}. First, we define the  
adjacency matrix $A$ by $A_{ij}=1$ if node $j$ 
points to node $i$ (if there is a link from the Wikipedia 
article $j$ to the article $i$) and $A_{ij}=0$ otherwise. 
Then the stochastic matrix $S$ of Markov transitions 
if defined by $S_{ij}=A_{ij}/\sum_{l} A_{lj}$ for columns $j$ 
with $\sum_{l} A_{lj}>0$. 
For dangling nodes $j$, having no outgoing links, i.~e. with 
$\sum_{l} A_{lj}=0$, we simply define $S_{ij}=1/N$ for all values of $i$. 
The columns of $S$ are sum normalized, i.~e. $\sum_l S_{lj}=1$, and 
conserve the total probability when the matrix $S$ is multiplied to an 
arbitrary probability vector $P$, 
i.~e. $\sum_i \tilde P(i)=\sum_i P(i)=1$ if $\tilde P=SP$. 

Finally, the Google matrix elements are defined as 
\begin{equation}
  G_{ij} = \alpha S_{ij} + (1-\alpha) / N
\label{eq_gmatrix} 
\end{equation}
where $0.5\leq \alpha <1$ is the damping factor for which we choose 
the usual standard  value $\alpha=0.85$ \cite{meyer,rmp2015}. 
The columns of $G$ are also column sum normalized and it also conserves 
the probability in the same way as the matrix $S$. 

Physically, this matrix describes a stochastic process 
where a random surfer jumps over the network. 
With a probability $\alpha$ the surfer jumps randomly from his actual 
page $j$ to a random page page $i$ among the pages with existing 
links $j\to i$ 
and with a complementary probability $\left(1-\alpha\right)$  he jumps to 
an arbitrary random node of the network. For dangling nodes he jumps 
immediately to an arbitrary random node. 
The damping factor allows to connect all isolated communities and avoids 
that the random surfer is trapped inside a small finite subset of nodes 
with no links going outside to this subset. 

\subsection{PageRank, CheiRank}

Due to the damping factor the stochastic process defined by $G$ converges, 
according to the Perron-Frobenius theorem, in the long time limit to 
a stationary probability $P=\lim_{t\to \infty} G^t P_0$ 
where $P_0$ is an arbitrary initial probability vector (with real 
values, $P_0(i)\ge 0$, $\sum_i P_0(i)=1$ and the same holds also for 
the stationary limit $P(i)$). 
The stationary vector $P$, in the following also called {\em PageRank} 
vector, satisfies the eigenvalue equation $GP=\lambda P=P$ with 
the maximal eigenvalue $\lambda=1$ of $G$. 
One can show from (\ref{eq_gmatrix}) 
that all other (real or complex) eigenvalues $\lambda$ of $G$ satisfy 
the inequality $|\lambda|\le \alpha$ 
providing a spectral gap $1-\alpha$ ensuring the exponential convergence 
to the steady state vector and the iteration $G^t P_0$ indeed provides 
a reliable numerical method to compute the PageRank. 

The value of the PageRank component $P(i)$ represents the {\em importance} 
of the node $i$ which is essentially proportional to 
the number of all ingoing links 
$j\to i$ \cite{meyer} but also weighted with the importance $P(j)$ 
of the source nodes $j$. It is also useful to determine the rank index 
$K_M(i)=1,2,3,\ldots$ of a node $i$ by ordering the nodes $i$ with 
decreasing values of $P(i)$. This index is also called PageRank index 
such that $K_M(i)=1$ for maximal $P(i)$, $K_M(i)=N$ for minimal $P(i)$ 
and more generally $K_M(i)< K_M(j)$ corresponding to $P(i)> P(j)$. 
The Google search engine actually uses this PageRank index to select the 
presentation order of search results which typically result in rather 
long lists of web pages. 

Following \cite{cheirank,wikizzs}, we can also consider the network 
obtained by the inversion of all the directions of links of the 
original network 
with a resulting Google matrix noted $G^*$ (which is different from the 
simple transpose $G^T$ due to different column-sum normalizations and 
different sets of dangling nodes). The PageRank vector associated to $G^*$ 
is called {\em CheiRank} vector and noted $P^*$ with a {\em CheiRank} ordering 
index $K_M^*(i)$. The value of $P^*(i)$ is typically proportional to 
the number of outgoing links $i\to j$ with some weight factors being 
$P^*(j)$. 

It is also useful to introduce a 2DRank index $K_2$ which orders nodes on the 
PageRank-CheiRank plane by order of their appearance in a square 
of increasing size $K_M=K_M^*$ starting from $K_M=K_M^*=1$ 
(see details in \cite{wikizzs}). The coarse grained density of the 
distribution of nodes in the $K_M$-$K_M^*$ plane, eventually in log-scale, 
provides also a useful graphical presentation of the network nodes; see also 
Figures\ref{fig1} and \ref{fig2} below. 

In this work, we note by $K_M$ ($K_M^*$) the PageRank index for the full 
(inverted) network while $K$ ($K^*$) denotes a reduced limited index 
for the group(s) of 40 nodes corresponding to Tables~\ref{table2} 
and \ref{table3}, i.~e. with 
values $K=1,\ldots,40$ ($K^*=1,\ldots,40$). 
These local group indices $K$ and $K^*$ can be computed by different methods, 
e.g. by direct extraction from $K_M$ and $K_M^*$ by attributing the values 
of $K$ ($K^*$) with 
increasing values of $K_M$ ($K_M^*$), e.g. $K=1$ for the subset node 
with minimal $K_M$ value, $K=2$ for the subset node with the second 
minimal $K_M$ value etc. We used however a slightly different method 
which consists simply of reducing 
the PageRank vector to the subset, with new index values being 
$1,\ldots,40$, and then computing $K$ by ordering the components of this 
reduced (or projected) PageRank vector (and similarly for $K^*$ 
using a reduced CheiRank vector). 
A third, more complicated method is to compute the projected PageRank 
from the reduced Google (see next section). 
Table~\ref{table3} summarizes these 
local indices $K$ and $K^*$ for the 8 editions and their associated 
subsets of 40 nodes.

\subsection{Reduced Google matrix}
\label{subsec3.3}

In \cite{politwiki} a method and algorithm (REGOMAX) 
was introduced to define and compute a {\em reduced Google matrix} 
for a selected subset of $\Nr$ nodes (with $N_r\sim 10^2$-$10^3$ being 
typically of modest size) taking into account both direct 
links between two nodes of this subset and also indirect links between 
two such nodes using pathways along nodes outside this subset. 
We note that the contributions of indirect links are very 
important and their omission may lead to erroneous results as it was 
demonstrated in \cite{eomwiki24} for a directed network of historical 
figures of Wikipedia previously studied in \cite{aragon}.

For this approach it is convenient to write the Google matrix $G$ and the 
PageRank vector $P$ of  
the global network as
\begin{equation}
\label{eq_Gblock}
G=
\left(\begin{array}{cc}
\Grr & \Grs \\
\Gsr & \Gss
\end{array}\right)
\quad,\quad
P=\left(\begin{array}{c}
\PR \\
\PS
\end{array}\right)
\end{equation}
where the label ``$\rm r$'' refers to the nodes of the reduced network 
of $\Nr$ subset nodes, and ``$\rm s$'' to the other $\Ns=N-\Nr$ nodes 
which form the complementary network acting as an effective 
``scattering network''. The PageRank eigenvalue equation $GP=P$ implies 
\cite{politwiki} that the projected 
PageRank $\PR$ can be computed from 
\begin{equation}\label{eq:GRPP}
	G_{\rm R}\PR=\PR 
\end{equation}
where the {\em reduced Gooogle matrix} of size $\Nr\times\Nr$ is given by
\begin{equation}
\label{eq_Geff1}
\GR=\Grr+\Grs({\bf 1}-\Gss)^{-1} \Gsr=\Grr + \Gpr + \Gqr\ .
\end{equation}
This matrix is also a stochastic matrix with columns being sum normalized. 
For practical reasons, we renormalize $\PR$ such that $\sum_j\PR(j)=1$ 
where the sum runs over all nodes of the small subset. In (\ref{eq_Geff1}), 
the first term $\Grr$ accounts for all direct links between two nodes 
$A$ and $C$ in the subset and the second term represents all indirect links 
between these two nodes using a chain of links
from $A$ to $B_1$, then from $B_1$ to $B_2$, $\ldots$, and then from 
$B_m$ to $C$ where the intermediate 
nodes $B_1,\ldots,B_m$ belong to the complementary scattering 
network of $\Ns$ nodes. Such a pathway corresponds to the term 
$\Grs \Gss^m \Gsr$ obtained by expanding the above matrix inverse in 
a geometric series over such terms. We note that the matrix $\Gss$ 
has a leading eigenvalue $\lambda_c$ close to 1 but smaller than 1 
(if $\Nr>0$) such 
that the matrix inverse is well defined. The second term 
can be furthermore decomposed \cite{politwiki} in two contributions 
$\Gpr + \Gqr$ where $\Gpr$ is a rank-1 matrix obtained by extracting 
from the matrix inverse the contribution of the leading eigenvector 
of $\Gss$ (which is rather close to $\PS$) and $\Gqr$ is the remaining 
contribution which can be numerically efficiently computed by a rapid 
convergent geometric series over a certain specific matrix obtained from 
$\Gss$ by a projection on the space bi-orthogonal to the leading eigenvector 
of $\Gss$ (see \cite{politwiki} for details). 
Initially, this decomposition was introduced to find an efficient numerical 
algorithm to $\GR$ but it turns out that it also useful in terms of 
interpretation. 
The term $\Gpr$, while having a dominant weight, has 
a very simple structure with nearly 
identical columns being close to $\PR$ 
(all columns are proportional with factors close to 1). This term is 
essentially determined by the leading eigenvector of $\Gss$ (see 
\cite{politwiki} for an explicit formula). 

Due to the simple structure of $\Gpr$ it is the other matrix $\Gqr$, 
smaller in numerical weight, which 
represents the most nontrivial information
related to indirect hidden transitions. 
We also define the matrix $\Gqrnd$ which is obtained from 
$\Gqr$ with its diagonal elements being replaced by zero. 
We note that each component can be characterized
by its weight being $\WR$, $\Wpr$, $\Wrr$, $\Wqr$ ($\Wqrnd$) for
$\GR$, $\Gpr$, $\Grr$, $\Gqr$ ($\Gqrnd$) respectively and which is 
defined as the sum of all matrix
elements divided by its size $\Nr$. (Since $\GR$ is also column sum 
normalized we always have $\WR=1$.)
Studies for examples of reduced Google matrices associated to various 
directed networks can be found in 
\cite{politwiki,wrwu2017,wikibank,wtn3,grmetacore}.

We note that the first equivalence relation in (\ref{eq_Geff1})
is similar to the Schur complement in linear algebra (see e.g. \cite{schur}).
Issai Schur introduced the Schur complement in 1917 (see history 
in \cite{schur}) and later it found a variety of applications \cite{schur,meyer2}.
However, the expansion in the form of three matrix components,
given by the second equivalence relation in (\ref{eq_Geff1}),
with the physical sense of each component was introduced only in \cite{politwiki}.

In this work, we apply the reduce Google matrix approach 
to the group(s) of 40 Wikipedia articles shown in Tables~\ref{table2} and 
\ref{table3} and introduced in the last section. 
For all these cases we computed the reduced Google matrix and its different 
components, as well as the reduced PageRank vector $\PR$ and the group local 
PageRank index $K$ (with values $K=1,\ldots,40$). The same can also be done 
for the inverted network with Google matrix $G^*$. However, in this 
work we present  only  results for the reduced Google matrix associated 
to $G$ and not to $G^*$ 
(except for the local index $K^*$ which can also be obtained more 
directly by extraction from $K_M^*$ or by ordering $\PR^*$ which is defined 
in a similar way as $\PR$). 

We present the positions of 40 articles in the PageRank-CheiRank plane for all 8 editions
in Figures~\ref{fig1} and \ref{fig2}; the reduced Google matrix and its components in
Figures~\ref{fig3}, \ref{fig4}, \ref{fig5}, \ref{fig6}, \ref{fig7} 
and related network diagrams in
Figures~\ref{fig8}, \ref{fig9}, \ref{fig10}, \ref{fig11}, \ref{fig12}, 
\ref{fig13}, \ref{fig14}, \ref{fig15}
for all 8 editions.

\subsection{Inter edition correlator quantities}
\label{subsec3.4}

It is interesting to compare the different quantities we compute, such 
as $\PR$, $K$, $\GR$ etc., between the different Wikipedia editions 
of Table~\ref{table1}. For this we consider different types of correlators 
for which we give a brief review below. 

For example for two given data sets $X_i$ and $Y_i$ of size $n_s$ with 
$i=1,\ldots,n_s$ the Pearson correlation coefficient is defined as 
\cite{wikicorr1}
\begin{equation}
\rho_{XY}=
\frac{\langle (X-\langle X\rangle)(Y-\langle Y\rangle)\rangle}
{\sigma_X\sigma_Y}=
\frac{\langle XY\rangle-\langle X\rangle\langle Y\rangle}
{\sigma_X\sigma_Y}
\label{defcorr1}
\end{equation}
where $\langle f(X,Y)\rangle=\sum_i f(X_i,Y_i)/n_s$ is the average over 
an arbitrary function $f(X,Y)$ and 
$\sigma_X=\sqrt{\langle X^2\rangle-\langle X\rangle^2}$, 
$\sigma_Y=\sqrt{\langle Y^2\rangle-\langle Y\rangle^2}$
is the standard deviation of $X$ and $Y$ respectively. For 
$\rho_{XY}>0$ ($\rho_{XY}<0$) one can state (for a given index $i$) 
that it is more likely 
that $Y_i>\langle Y\rangle$ ($Y_i<\langle Y\rangle$) if $X_i>\langle X\rangle$.
In other words deviations from the average values for $X$ and $Y$ 
have probably the same (opposite) sign if $\rho_{XY}>0$ ($\rho_{XY}<0$). 

The Kendall rank correlation coefficient is defined as \cite{wikicorr2}
\begin{equation}
\tau_{XY}=\frac{2n_c}{n_p}-1
\label{defcorrKendall}
\end{equation}
where $n_p=n_s(n_s-1)/2$ is the number of all possible pairs 
$(X_i,Y_i)$, $(X_j,Y_j)$ with $i<j$ and $n_c$ is the number of 
{\em concordant pairs} such that 
either $X_i<X_j$ and $Y_i<Y_j$ or $X_i>X_j$ and $Y_i>Y_j$ (same sorting 
order between the pair). Here, we have $\tau_{XY}=1$ ($\tau_{XY}=-1$) 
if both data sets have the same (reverse) sorting order. 

In the (end of the) next section and in Figure~\ref{fig16}, 
we present and discuss results for both 
correlators $\rho_{q(i)q(j)}$ and $\tau_{q(i)q(j)}$ as 
$8\times 8$ matrices 
where $i,j=1,\ldots, 8$ represent index values for the 8 Wikipedia editions 
and $q(i)$ a data set/quantity computed for the given Wikipedia edition 
with index value $i$. 
Concerning the correlator (\ref{defcorr1}), we consider five 
quantities $q(i)$ being the reduced PageRank vector $\PR$ (all vector 
coefficients with $n_s=40$), the reduced Google matrix $\GR$ 
(all matrix elements with $n_s=40^2$), 
the religion or society sub-block of $\Grr+\Gqrnd$ (all sub-block 
matrix elements with $n_s=17^2$ or $n_s=23^2$ respectively) and also the 
local PageRank index $K$ (all $K$ values with $n_s=40$). 
For $K$ we also compute the Kendall rank correlator 
(\ref{defcorrKendall}) which is actually equivalent to 
the Kendall rank correlator for $\PR$ since the local ranking index 
$K$ is obtained by ordering the values of $\PR$, i.~e. two $K$ vectors 
for two different editions have the exact same number of concordant pairs as 
the related two $\PR$ vectors. We mention that mathematically 
both correlator quantities can take in theory values between 
$-1$ (perfect anti-correlation) and $+1$ (perfect correlation) while 
values close to 0 indicate small or absent correlations. However, 
for the 6 correlator quantities mentioned above only positive correlator 
values appear (actually with minimal values above $0.33$; 
see Figure~\ref{fig16} for more details).

\section{Results }
\label{sec4}

\subsection{Article location on PageRank-CheiRank plane}


\begin{figure}
\centering
\includegraphics[width=0.8\textwidth]{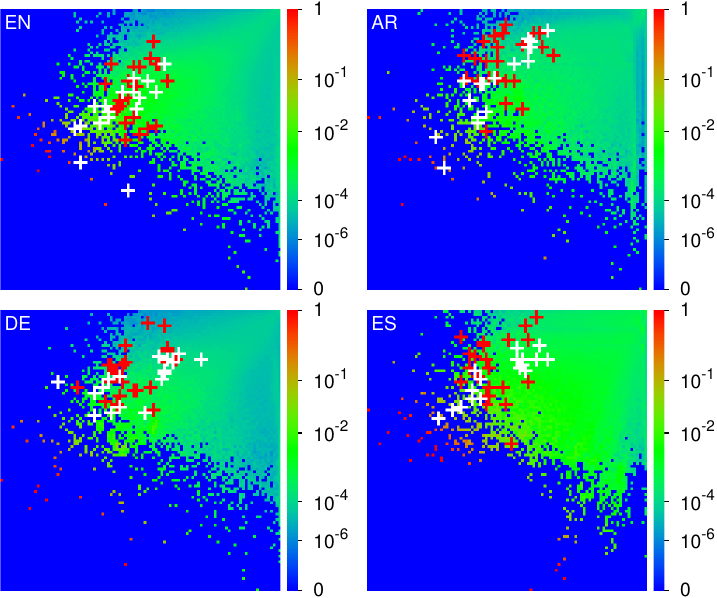}
\caption{\label{fig1} 
Density of nodes  $W(K_{\rm M},K_{\rm M}^*)$ 
on PageRank-CheiRank plane $(K_{\rm M},K_{\rm M}^*)$
averaged over $100\times100$ 
logarithmically equidistant grids for 
$0 \leq \ln K_{\rm M}, \ln K_{\rm M}^* \leq \ln N$ 
($1 \leq  K_{\rm M},K_{\rm M}^* \leq N $) for the four Wikipedia editions EN (top-left), 
AR (top-right), DE (bottom-left) and ES (bottom-right); 
the values of node number $N$ for each edition are given in 
Table~\ref{table1}; 
the density is averaged over all nodes inside each cell of the grid,
the normalization condition is 
$\sum_{K_{\rm M},K_{\rm M}^*}W(K_{\rm M},K_{\rm M}^*)=1$.
Color varies from blue at zero value to red at maximal density value.
The numbers at the color bar correspond to $W(K_{\rm M},K_{\rm M}^*)/W_{\rm cut}$ 
where $W_{\rm cut}=\max W(K_{\rm M},K_{\rm M}^*)/16$ 
and values of $W(K_{\rm M},K_{\rm M}^*)>W_{\rm cut}$
have been saturated to $W_{\rm cut}$. The non-linear color scale 
(corresponding to $x^8$ if $x\in[0,1]$ represents 
the linear scale of the visible color bar) and the saturation at 
$W_{\rm max}/16$ have been chosen in order to increase the visibility of 
low density values. The $x$-axis corresponds to $\ln K_{\rm M}$ 
and the $y$-axis to $\ln K_{\rm M}^*$ with $K_{\rm M}$ ($K_{\rm M}^*$) 
being the global PageRank (CheiRank) index for the Wikipedia network of 
the corresponding edition. 
The red (white) crosses mark the positions of the 23 society nodes 
with $K_g\le 23$ (17 religion nodes with $K_g\ge 24$) 
of Tables~\ref{table2} and \ref{table3}. 
}
\end{figure}

We first discuss some results for the local ranking indices 
$K$ and $K^*$ given in Table~\ref{table3} for the 8 edition cases 
and the 40 edition specific selected articles. 
Concerning EN-Wikipedia, we note that the top 6 PageRank positions 
in the subset of Table~\ref{table2} are taken by religions with 
Catholic Church at $K=1$ while the first society concept Law only appears 
at $K=7$. This tendency is preserved in the other 7 editions
where religions still take the top 3-4 PageRank positions
while such a society concept as Liberalism only appears at $K=5$ for the 
AR, ES, FR, IT editions. There are two exceptions being 
Nazism at $K=2$ for DE, that is clearly linked to German history, 
and surprisingly also for ZH where 
the top society concept is Law at $K=3$
being significantly above Communism at $K=13$.
For EN the top society concept is also Law at $K=7$
and for RU it is Economy at $K=7$.

Among religions the top $K=1$ position
is taken by Catholic Church for EN, DE, ES, IT 
and Christianity is at $K=1$ for FR, ZH 
(Buddhism is at $K=2$ for ZH).
Somewhat surprisingly for RU Islam is at the top PageRank
position $K=1$ and Christianity is at $K=2$. This 
is probably related to a significant Muslim population in Russia
that is however significantly smaller than its christian population.
Islam is the top religion for the AR-Wikipedia edition which appears to be 
rather natural. However, here Christianity and Catholic Church appear 
at $K=2$ and $K=3$ respectively. 

Concerning the CheiRank index $K^*$, that characterizes the communicative 
properties of a node, the situation is more mixed. Thus we have at $K^*=1$:
Jainism for EN; Christianity for AR; Islam for DE,
Anarchism for ES, FR; Catholic Church for IT; Monarchy for RU;
Communism for ZH. We attribute such a mixing to
stronger fluctuations of outgoing links as discussed in \cite{rmp2015}.

\begin{figure}
\centering
\includegraphics[width=0.8\textwidth]{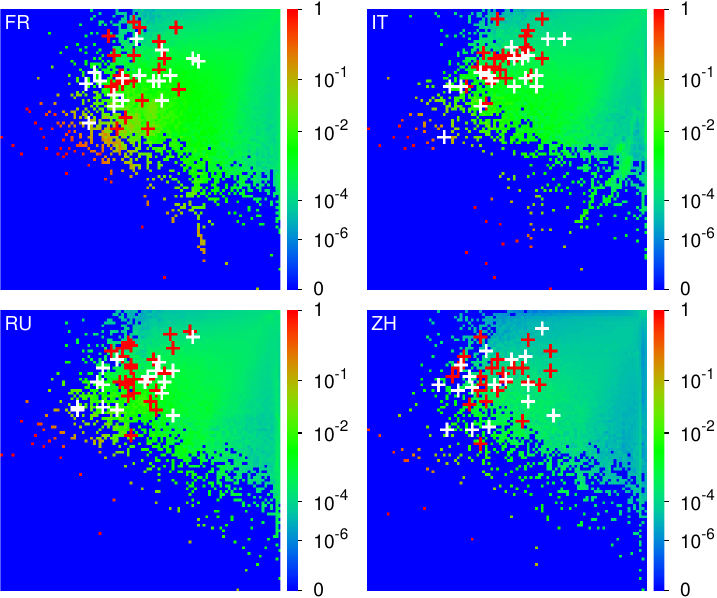}
\caption{\label{fig2} 
As Figure~\ref{fig1} but for the four Wikipedia editions FR (top-left), 
IT (top-right), RU (bottom-left) and ZH (bottom-right).
}
\end{figure}

Concerning the global ranking indices $K_M$ and $K_M^*$ for the full network, 
we show the density of their distribution in the 
$\ln(K_{\rm M})$-$\ln(K_{\rm M}^*)$ plane in Figure~\ref{fig1} 
for EN, AR, DE, ES and Figure~\ref{fig2} for FR, IT, RU, ZH. The overall 
density structure is rather comparable for all 8 editions with 
a certain asymmetry between both axis. However, the precise 
top $K_M$ and $K_M^*$ positions (isolated small red or green squares) are very 
specific for each edition. 

Furthermore, the locations of the 40 edition specific selected articles 
of Tables~\ref{table2} and \ref{table3} 
are also shown in Figures~\ref{fig1} and \ref{fig2}. 
We see that for all editions religion nodes (white crosses corresponding 
to $24\le K_g\le 40$) have typically higher PageRank positions 
(lower $K_M$ values) as compared to social concepts (red crosses 
corresponding to $1\le K_g\le 23$), i.~e. typically 
the top religion nodes appear to be more important than the top society nodes. 
In particular, for the EN-edition the top religion 
PageRank index is about 6 times smaller as compared to the top social 
concept PageRank index as can be seen from Table~\ref{table2}. 
For the other editions, we have similar ratios of $K_M$ between 
the top religion and the top society nodes which can be seen from the 
difference of the horizontal positions between the first white cross 
and the first red cross (difference of $\ln K_M$) in the different panels of 
Figures~\ref{fig1} and \ref{fig2}.

\subsection{Matrix components of REGOMAX algorithm}

We have computed the different reduced Google matrix components 
(as described in Subsection \ref{subsec3.3}) 
for the 8 Wikipedia edition and using for each edition the edition 
specific group of 40 articles given in Tables~\ref{table2} and \ref{table3} 
(as explained in Subsection \ref{subsec2.2}). 
The presentation order of the nodes in the groups is given by the index 
$K_g$ which was obtained by separately PageRank ordering the 23 nodes 
for society concepts and the 17 nodes for religions 
(or branches of religions) for the EN-Wikipedia edition. However, for 
the other editions there are different $K$ and $K^*$ indices but 
for practical reasons we keep the same initial node order $K_g$ (obtained 
from EN-Wikipedia) when presenting the different matrix components in 
the figures below. 

We remind that $\GR$ is according to (\ref{eq_Geff1}) given by 
the sum $\Grr + \Gpr + \Gqr$ where $\Grr$ represents the direct links, 
$\Gpr+\Gqr$ indirect links with $\Gpr$ being a rank-1 matrix 
(with columns being rather close to the projected PageRank $\PR$) 
taking into account the contributions of the leading eigenvector 
in the complementary scattering space (see \cite{politwiki} for details) 
and $\Gqr$ is obtained from the other contributions and containing 
interesting nontrivial information about indirect links. 
Numerically, $\Gpr$ is quite dominant in 
this sum with relative weights $\Wpr=0.91$-$0.97$ 
depending on the edition but it has a very simple structure. 
The interesting properties of the reduced Google matrix are contained 
in the matrix $\Grr+\Gqrnd$ being the sum of $\Grr$ and $\Gqr$ (with 
diagonal elements of the latter replaced by 0). 
For two editions (EN and ZH), we present results for the four 
matrices $\GR$, $\Gpr$, $\Grr$ and $\Grr+\Gqrnd$ while for the other six 
editions we limit ourselves to $\GR$ and $\Grr+\Gqrnd$. 

Below we will focus our discussion on the 
most interesting case $\Grr+\Gqrnd$. To simplify this discussion, 
we also introduce for this matrix the sub-blocks $A$ 
(left top diagonal society block), 
$B$ (right top block with links from religion to society), 
$C$ (left bottom block with links from society to religion) 
and $D$ (bottom right diagonal religion block). 
In particular, we determine the strongest matrix element for each block, 
the sum of matrix elements per block and 
the ratios $R(C,A)$, $R(D,B$, $R(D,A)$ for these sums. 
In the next subsection, we will see that 
the matrix $\Grr + \Gqrnd$ can also be exploited to 
generate effective networks of friends and followers. 

\begin{figure}
\centering
\includegraphics[width=0.8\textwidth]{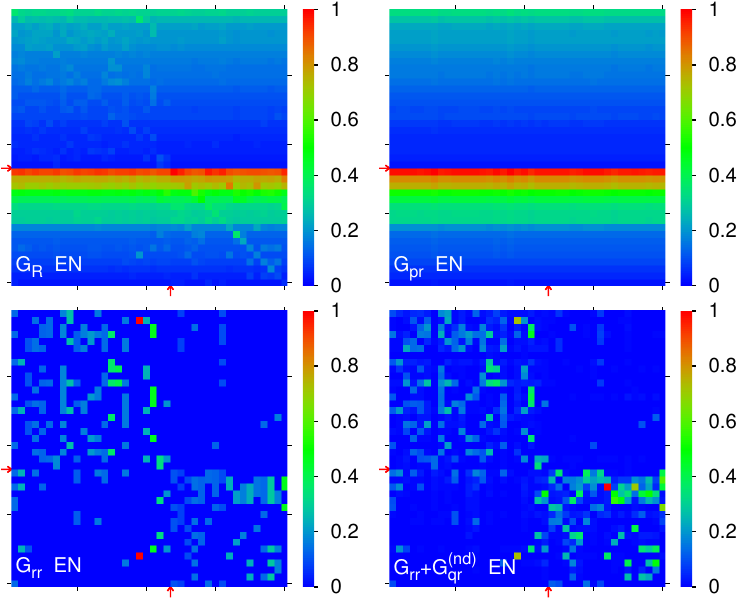}
\caption{\label{fig3} 
Color density plots of the matrix components $\GR, \Gpr, \Grr, \Grr+\Gqrnd$
for the group of Table~\ref{table2} and  Wikipedia EN edition; 
the $y$-axis corresponds to the first (row) index (increasing values 
of $K_g$ from top to down) and the $x$-axis corresponds to the second 
(column) index of the matrix (increasing values of $K_g$ from left to right). 
The outside tics indicate multiples of 10 of $K_g$. 
The red arrows indicate the separation between society nodes ($K_g\le 23$) 
and religion nodes ($K_g\ge 24$) in both axis. 
The numbers in the color bar 
correspond to $g/g_{\rm max}$ with $g$ being the 
value of the matrix element and $g_{\rm max}$ being the maximum value.
For $\Gqrnd$ there are some small negative matrix elements corresponding 
to values $g/g_{\rm max}>-0.035$ ($g/g_{\rm max}>-0.038$ for other editions 
shown in other figures below) which are shown with a color very close 
to blue for zero values.
}
\end{figure}

In Figure~\ref{fig3}, we show color density plots for  
$\GR$, $\Gpr$, $\Grr$ and $\Grr+\Gqrnd$ for the edition EN 
with numerical weights being 
$\WR=1$, $\Wrr=0.0144$, $\Wpr = 0.9661$, $\Wqr=0.0194$ ($\Wrr+\Wqrnd=0.0280$). 
The color plot of $\Gpr$ is essentially composed of rows of equal color 
with matrix elements $\Gpr(i,j)\approx \PR(i)$ for all columns $j$. 
Here the sequence of row colors illustrates the separate 
PageRange order in the two blocks. 
The row with red color corresponds to the top PageRank node 
``Catholic Church'' with $K=1$, $K_g=24$, first row in the religion block,
and the rows below at $K_g=25,\ldots,29$ with orange or (strong) green 
color correspond to $K=2,\ldots,6$. 
The color plot of $\GR$ is similar in appearance, due to the strong weight 
of $\Gpr$, but with additional peaks at isolated positions with 
largest elements of $\Grr$ or $\Gqr$ (including some quite strong diagonal 
elements of $\Gqr$, especially in the religion block). 

For $\Grr+\Gqrnd$ the strongest matrix elements $\Grr(i,j)+\Gqrnd(i,j)$ for 
each block $A$, $B$, $C$ and $D$ (for links $i \leftarrow j$) correspond to: 
0.0126 (Education $\leftarrow$ Oligarchy, $A$);
0.0021 (Economy $\leftarrow$ Chinese folk religion, $B$); 
0.0125 (Shia Islam $\leftarrow$ Oligarchy, $C$); 
0.0172 (Islam $\leftarrow$ Sunni Islam, $D$). 
The last value appears as a sum of a direct link and also a stronger 
indirect link from Sunni Islam to Islam while the two links 
from Oligarchy to Education or Shia Islam result from direct links 
also clearly visible in $\Grr$. 

We also observe that the two 
diagonal blocks $A$ and $D$ seem rather decoupled with 
significantly smaller links in the off-diagonal blocks $B$ and $C$ 
which is also confirmed by the sum ratios 
$R(C,A) = 0.2778$, $R(B,D) = 0.0996$, $R(D,A) = 0.8998$.
The value of  $R(D,A)$ is above to the ratio of areas
$(17/23)^2 \approx  0.55$ showing that
the transitions inside the religion block $D$ are more intense
comparing to the society concepts block $A$,
this difference  is also related  to the particularly 
strong maximal element in the $D$ block (see above). 
The decoupling between the two blocks, which is also be confirmed 
for the other 7 Wikipedia editions discussed below, 
seems to be surprising in view
of the important historical role played by religions
in society formation. However, on the other side
there is a well known statement from the Bible:
{\it Render unto Caesar the things that are Caesar's, and unto God the things that are God's}
(Bible Matthew 22:21 \cite{bible})
that  may be partially at the origin of this result.
Also in certain countries, e.g. France,
the separation between the state and religions is fixed by law.

\begin{figure}
\centering
\includegraphics[width=0.8\textwidth]{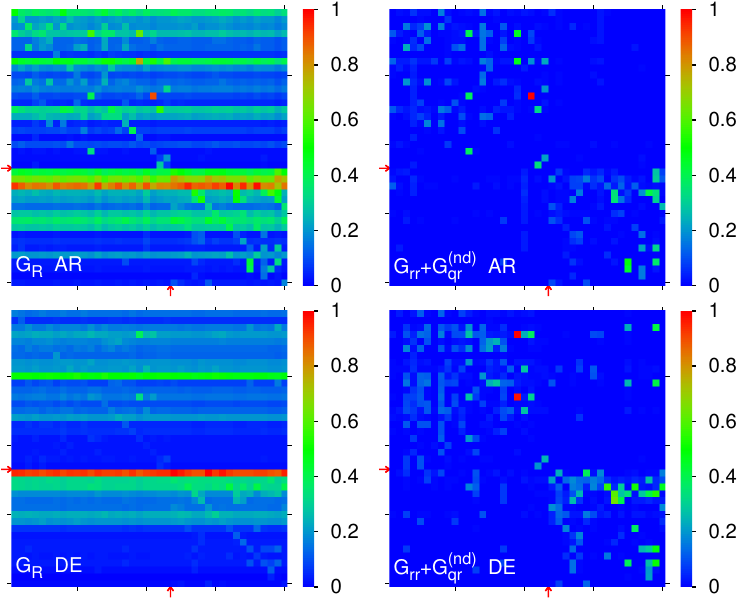}
\caption{\label{fig4} 
Color density plots of the matrix components $\GR, \Grr+\Gqrnd$ for 
the edition specific group/network (see also Table~\ref{table3}) 
of AR and DE. The technical details 
for the color plot presentation are exactly as in Figure~\ref{fig3}.
}
\end{figure}

Figure~\ref{fig4} shows the matrices $\GR$ and $\Grr + \Gqrnd$ 
for the two editions $AR$ and $DE$. For AR we have the weights 
$\Wpr = 0.9153, \Wrr = 0.0452, \Wqr = 0.03936$
with higher weights of $\Grr$ and $\Gqr$ as compared to EN. 
For AR the strongest matrix elements of $\Grr+\Gqrnd$ per block correspond to: 
0.1067 (Monarchy $\leftarrow$ Autocracy, $A$);
0.0063 (Law $\leftarrow$ Catholic Church, $B$);
0.0132 (Islam $\leftarrow$ Law, $C$);
0.0524 (Taoism $\leftarrow$ Confucianism, $D$).
The first element of this list is related to the fact 
that several islamic countries are monarchies. 
The sum ratios are $R(C,A) = 0.1311$, $R(B,D) = 0.1053$, $R(D,A) = 0.7031$.
Here in the panel of $\GR$ one sees a strong red row at $K_g=26$ 
for the top PageRank node Islam. 

In a similar way, we obtain for DE 
$\Wpr = 0.9582, \Wrr = 0.04200, \Wqr = 0.0217$ and here 
the strongest matrix elements of $\Grr+\Gqrnd$ per block correspond to: 
0.0327 (Democracy $\leftarrow$ Oligarchy, $A$);
0.0137 (Communism $\leftarrow$ Chinese folk religion, $B$);
0.0058 (Islam $\leftarrow$ Republic, $C$);
0.0209 (Hinduism $\leftarrow$ Jainism, $D$).
The sum ratios are $R(C,A) = 0.1364$, $R(B,D) = 0.1847$, $R(D,A) = 0.8417$. 
As for EN there is a red row for $\GR$ at $K_g=24$ for Catholic Church. 

\begin{figure}
\centering
\includegraphics[width=0.8\textwidth]{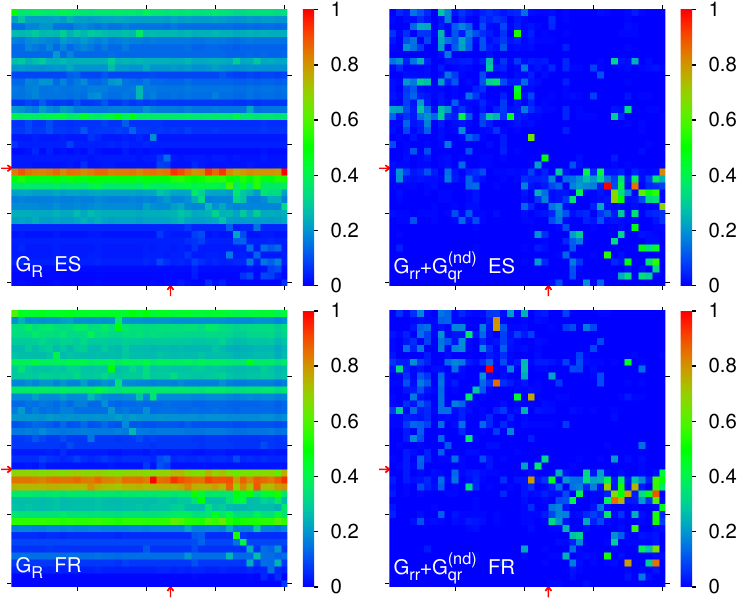}
\caption{\label{fig5} 
Color density plots of the matrix components $\GR, \Grr+\Gqrnd$ for 
the edition specific group/network (see also Table~\ref{table3}) 
of ES and FR. The technical details 
for the color plot presentation are exactly as in Figure~\ref{fig3}.
}
\end{figure}

For the ES edition (top panels of Figure~\ref{fig5}) 
we find $\Wpr = 0.9372$,
$\Wrr =  0.0269$, $\Wqr = 0.0358$ and 
the strongest matrix elements of $\Grr+\Gqrnd$ per block correspond to: 
0.0222 (Oligarchy $\leftarrow$ Autocracy, $A$);
0.0117 (Society $\leftarrow$ Confucianism, $B$);
0.0082 (Buddhism $\leftarrow$ Materialism, $C$);
0.0346 (Islam $\leftarrow$ Sunni Islam, $D$).
Here we have in the $D$ block a second very strong matrix element 
with value 0.0296 (Islam $\leftarrow$ Shia Islam). 
The sum ratios are given by 
$R(C,A) = 0.2170$, $R(B,D) = 0.1403$, $R(D,A) = 1.0227$.
As for EN and DE there is a red row for $\GR$ at $K_g=24$ for Catholic Church. 

For FR (bottom panels of Figure~\ref{fig5}) 
we have $\Wpr = 0.9572$, $\Wrr =  0.0197$, $\Wqr = 0.0229$ and 
the strongest matrix elements of $\Grr+\Gqrnd$ per block correspond to: 
0.0222 (Culture $\leftarrow$ Society, $A$);
0.0101 (Politics $\leftarrow$ Confucianism, $B$);
0.0178 (Christianity $\leftarrow$ Autocracy, $C$);
0.0198 (Buddhism $\leftarrow$ Chinese folk religion, $D$) with 
additional strong matrix elements 
0.0191 (Capitalism $\leftarrow$  Economy, $A$), 
0.0179 (Education $\leftarrow$  Economy, $A$), 
0.0176 (Communism $\leftarrow$  Economy, $A$),
0.0189 (Taoism $\leftarrow$ Chinese folk religion, $B$) and
0.0170 (Monarchy $\leftarrow$ Autocracy, $A$). 
We attribute the last link and the top $C$ link from Autocracy 
to Christianity to the 
{\em bapt\^eme de Clovis} when the king of France Clovis I
accepted the Christian religion around the year 500.
The sum ratios are 
$R(C,A) = 0.1894$, $R(B,D) = 0.1395$, $R(D,A) = 1.1519$.
There is a red row for $\GR$ at $K_g=25$, $K=1$ for Christianity along 
with strong orange rows at $K_g=26$, $K=2$ (Islam) and 
$K_g=24$, $K=3$ (Catholic Church). 

\begin{figure}
\centering
\includegraphics[width=0.8\textwidth]{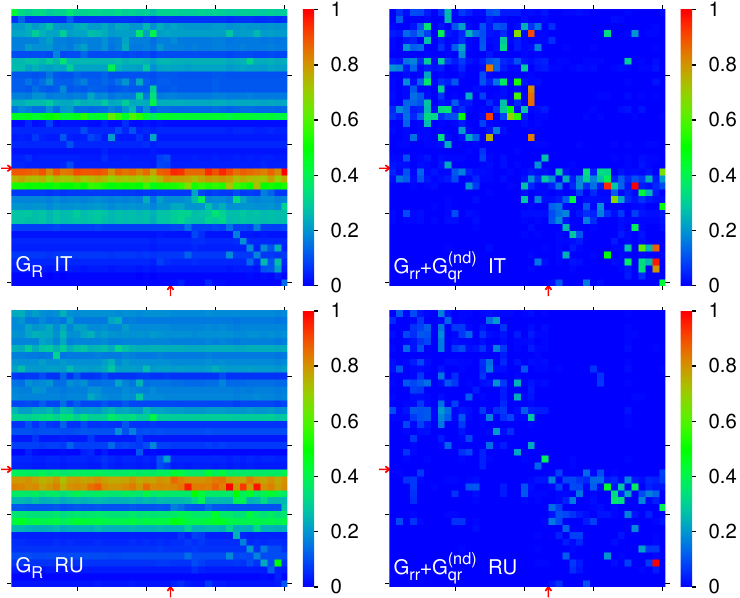}
\caption{\label{fig6} 
Color density plots of the matrix components $\GR, \Grr+\Gqrnd$ for 
the edition specific group/network (see also Table~\ref{table3})
of IT and RU. The technical details 
for the color plot presentation are exactly as in Figure~\ref{fig3}.
}
\end{figure}

For the IT edition (top panels of Figure~\ref{fig6}) the matrix weights 
are $\Wpr = 0.9393$, $\Wrr = 0.0297$, $\Wqr = 0.0308$ and 
the strongest matrix elements of $\Grr+\Gqrnd$ per block correspond to: 
0.0298 (Economy $\leftarrow$ Society, $A$);
0.0108 (Democracy $\leftarrow$ Sunni Islam, $B$);
0.0095 (Christianity $\leftarrow$ Idealism, $C$);
0.0326 (Islam $\leftarrow$ Shia Islam, $D$). 
Furthermore there are significant numbers of additional strong matrix 
elements in the $D$-block:
0.0313 (Taoism $\leftarrow$ Chinese folk religion);
0.0307 (Islam $\leftarrow$ Sunni Islam);
0.0285 (Confucianism $\leftarrow$ Chinese folk religion);
0.0260 (Shinto $\leftarrow$ Chinese folk religion)
and also in the $A$-block:
0.0292 (Democracy $\leftarrow$ Autocracy);
0.0280 (Fascism $\leftarrow$ Autocracy);
0.0278 (Monarchy $\leftarrow$ Autocracy);
0.0273 (Oligarchy $\leftarrow$ Autocracy);
0.0272 (Culture $\leftarrow$ Society);
0.0266 (Republic $\leftarrow$ Autocracy), the 
latter probably being related to Italian history. 
The sum ratios are given by 
$R(C,A) = 0.1372$, $R(B,D) = 0.1159$, $R(D,A) = 0.6477$ and there 
are two strongs rows for $\GR$ being red ($K_g=24$, $K=1$, Catholic Church) 
and orange ($K_g=25$, $K=2$, Christianity). 

For RU (bottom panels of Figure~\ref{fig6}) we have 
$\Wpr = 0.9515$, $\Wrr = 0.0217$, $\Wqr = 0.0267$ and 
the strongest matrix elements of $\Grr+\Gqrnd$ per block correspond to: 
0.0223 (Materialism $\leftarrow$ Idealism, $A$);
0.0043 (Capitalism $\leftarrow$ Protestantism, $B$);
0.0066 (Buddhism $\leftarrow$ Civilization, $C$);
0.0569 (Taoism $\leftarrow$ Chinese folk religion, $D$). 
Here we have the sum ratios 
$R(C,A) = 0.1519$, $R(B,D) = 0.0928$, $R(D,A) = 0.7268$ 
and in the $\GR$ panel we see two orange-red rows at 
$K_g=26$, $K=1$ (Islam) and, with slightly smaller values, at 
$K_g=25$, $K=2$ (Christianity). 

\begin{figure}
\centering
\includegraphics[width=0.8\textwidth]{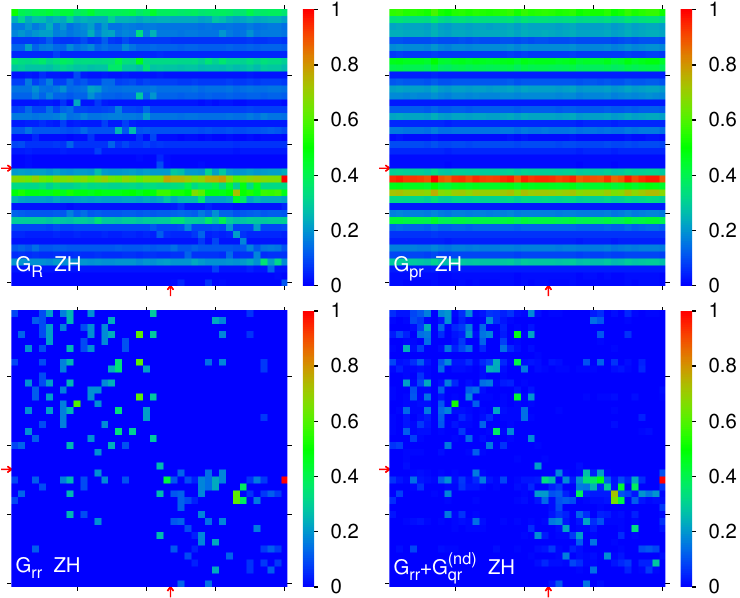}
\caption{\label{fig7} 
As Figure~\ref{fig3} but for the edition specific group/network of ZH.
}
\end{figure}

Finally, in Figure~\ref{fig7} we show the matrices 
for $\GR$, $\Gpr$, $\Grr$ and $\Grr+\Gqrnd$ of the ZH edition. Here the weights
are: $\Wpr = 0.9233, \Wrr = 0.0333, \Wqr = 0.0432$ ($\Wrr+\Wqrnd=0.0658$).
For ZH the strongest matrix elements of $\Grr+\Gqrnd$ per block correspond to: 
0.0341 (Fascism $\leftarrow$ Nazism, $A$);
0.0097 (Education $\leftarrow$ Judaism, $B$);
0.0219 (Christianity $\leftarrow$ Idealism, $C$);
0.0653 (Christianity $\leftarrow$ Oriental Orthodox Churches, $D$) and 
the block-sum ratios are $R(C,A) = 0.2639$, $R(B,D) = 0.1319$, 
$R(D,A) = 0.8458$. In the panel of $\Gpr$ we see a strong red row 
at $K_g=25$, $K=1$ (Christianity), which appears less pronounced 
(between orange and strong green) in the other panel of 
$\GR$ but mainly because 
the very strong maximal direct matrix element in the $D$-block of 
$\Grr$ (and of $\Grr+\Gqrnd$ corresponding to the link 
Christianity $\leftarrow$ Oriental Orthodox Churches) 
shifts the maximum value in the color plot defining the red color 
which reduces the color scale of other matrix elements. 
We note that the structure of the matrix $\Grr + \Gqrnd$ follows for ZH mainly
$\Grr$ for the strongest transitions. 

The  results of this subsection show
that for the matrices $\Grr + \Gqrnd$ of 
all 8 editions there are indeed multiple significant
transitions inside both blocks of society concepts and of religions. 
However, the transitions between these two blocks are, roughly by a factor 
5-10, smaller if we compare the sums of matrix elements of the 
off-diagonal block to the sums over the diagonal blocks. 
The ratio of transition strengths
of the two diagonal blocks is typically $R(D,B)$ somewhat 
larger than the ratio $(17/23)^2 \approx 0.55$ 
of block areas $D, B$.
This shows that transitions between religions
are on average a bit stronger than those between society concepts.


\subsection{Network structure inside social concepts and religions}
\label{subsec4.3}

In this section, we present effective network diagrams for friends 
and followers based on the information contained in the matrix $\Grr + \Gqrnd$ 
or more precisely in its two diagonal blocks $A$ for society concepts and 
$D$ for religions. Since, according to the results of the last subsection, 
these two blocks are rather well decoupled with only weak links between 
them, we will present separate network diagrams for each block. 
Network diagrams of (nearly) the same style, were for example used 
in \cite{politwiki} (for groups of political leaders in the 
Wikipedia network), \cite{wikibank} (for banks and countries in the 
Wikipedia network), and \cite{grmetacore} (for a specific fibrosis 
related protein group in the MetaCore network of proteins). 

However, for convenience, we present here the construction method of these 
network diagrams. Assume we have a small matrix $g$ with elements $g(i,j)$
being either a reduced Google matrix or one of its components (e.g. 
$\GR$, $\Grr$, $\Gqr$ or $\Grr+\Gqrnd$) or a certain sub-block of such 
a matrix (e.g. society or religion sub-blocks $A$ or $D$ of 
$\Grr+\Gqrnd$ shown in the previous section). 
First, we choose in the list of nodes (associated to this matrix or block) 
5 top nodes representing five different subgroups based on some 
categorization criteria (depending on the set of nodes and the context)
and we attribute each other node of this list to exactly one of the 5 
subgroups (based on some criteria and the context). 
In the following, we will use for these subgroups the 
notation {\em poles} as a synomym for ``center of interest'' which 
is a typical use of this expression in the French language. 
For each pole, we also define some presentation color. 

To construct the effective friend network (see below for the other 
case of follower networks), 
we draw first a main circle (thin gray line) and place the 5 top nodes 
uniformly on this circle with some label and the corresponding color. 
Then we select for each top node $j$ (also called level-1 nodes) 
the four strongest friends $i$ (level-2 nodes) 
with strongest outgoing links $j\to i$, i.~e. with 
largest matrix elements $g(i,j)$ in the same column $j$ of this matrix.
Each of these strongest friends, if not yet present in the diagram as another 
level-1 node, is placed on a smaller secondary 
circle around the top level-1 node associated to him and we draw thick 
black arrows from the level-1 nodes to their friends. It is possible 
that a new level-2 node appears as a friend of several initial level-1 
nodes. In this case, we try first to place this level-2 node on the circle 
of the level-1 node with same color (same pole) if possible, 
i.~e. if this level-2 
node is indeed a friend of the level-1 node of same color. Only if this is not 
possible, we place it on the circle of another level-1 node (first 
level-1 node of different color which has the given level-2 node as friend).
If a level-1 node has a friend which is already present in the diagram as 
another level-1 node, we simply draw a thick black arrow from the former 
to the latter and do not modify the position of the latter. 

The procedure is repeated for all (newly added) level-2 nodes with smaller 
circles around them on which we place their (up to) four strongest friends 
(level-3 nodes if newly added) and with the same rule for preferential 
placement on a circle of a parent node of same color. 
Now, we draw thin red arrows 
from the level-2 nodes to their friends. In case if such a friend is already 
present in the network (as level-1 or level-2 node), his position is not 
modified and we only draw the thin red arrow from his parent node to him. 
We also mention that only non-empty circles with at least one node 
on them are drawn; i.~e. 
if a given node has no newly added friends (i.~e. all his friend are 
already in the diagram), then we will not draw an empty circle around him. 

At this stage, we typically stop the procedure for simplicity. 
Even if we continue this procedure with level-4, level-5 nodes etc. 
the number of newly added nodes quickly decreases and when there is no 
newly added node the procedure converges to a stable final diagram. 
This can happen actually quite 
early so that their is typically no big difference in diagrams limited 
to level-3 nodes and those with higher level nodes. In particular, for 
the diagrams given below, the number of newly added level-3 nodes is typically 
already quite small (much smaller than the 
theoretical limit $5\times 4\times 4=80$)
also because the available set of nodes is limited from the very beginning, 
even significantly smaller than the theoretical level-3 limit. 
In some of the diagrams below 
there are even no newly added level-3 nodes (if absence of smallest level-3 
circles) and we have 
already convergence to a stable diagram at level-2, i.~e. all friends of 
level-2 nodes are already present in the diagram as former level-1 or level-2 
nodes. In case of convergence, the last stage does not add new nodes but 
it still adds arrows from the most recently added nodes in the previous stage 
to their friends (which are already present in the network diagram). 

The construction of follower network diagrams is essentially the same with 
two modifications: (i) at each level $k$ we select for each level-$k$ node 
$i$ (typically only $k=1,2$ in our case) the four strongest followers $j$ 
(as possible level-$(k+1)$ node if not yet present in the diagram) 
with strongest incoming links $i\leftarrow j$ defined by the largest 
matrix elements $g(i,j)$ in the same row $i$; 
(ii) arrows (thick black or thin red) are drawn with inverted directions 
from followers (level-$(k+1)$ nodes) to parents (level-$k$ nodes), 
e.g. there are 
some arrows from a circle node to its center node while in the 
friend diagrams we have arrows from the circle center to the outside nodes 
on the circle (note in case of multiple parent nodes or 
pre-existing friends or 
followers, we have typically a significant number of other type of arrows 
between different circles). 

In this work, we present figures for the network diagrams constructed 
from the two diagonal blocks $A$ for society related nodes and $D$ for 
religions of the matrix $\Grr+\Gqrnd$ for the 8 different Wikipedia 
editions. Since there are two friend and follower diagrams for each case, we 
have per edition four network diagrams presented in a figure with four panels. 
The subgroups or poles together with their respective 5 top nodes for 
both society and religion cases are given in Table~\ref{table2} 
(8th column) and this table also contains a two letter code for 
each node (5 th column) used as a node label in the diagrams. 

For society concepts, we choose the 5 top pole nodes 
Law, Society, Communism, Liberalism and Capitalism with 
respective node colors being olive, (dark) green, cyan, blue and indigo. 
We have tried to attribute the members of the poles based 
on the context and logical proximity to the top node, e.g. 
Education and Culture are attributed 
to the 2nd pole of Society; Ecology, Politics belong to the first pole 
of Law; Socialism and Anarchism are attributed to the 3rd pole for Communism.
In certain cases, this attribution is a bit arbitrary and other choices 
would have been possible. We also tried to assure that each pole has 
a certain minimum number of members. 

For the religion nodes, we choose the 5 top pole religions Christianity, 
Islam, Buddhism, Hinduism and Chinese folk religion (same colors as 
for the society top nodes in this order) and with its pole members being 
related to branches of religions or sub-religions. Here, 
we have attributed Judaism to the 1st Christianity pole with other members 
being Catholic Church, Protestantism and both nodes about Orthodox Church.

In the following, we will more precisely call this poles also ``initial 
poles'' in order to distinguish them from ``natural poles'' which may emerge 
naturally by certain clusters in a network diagram. Quite often 
natural poles and initial poles are very similar but in certain cases 
natural poles are composed of nodes from several initial poles.

Specifically, for the EN edition, whose network diagrams are shown 
in Figure~\ref{fig8}, we can identify in the friend society diagram 
the formation of 5 natural poles (which may slightly deviate from 
the initial poles) with main members being 
1T) Law, Politics, Monarchy, Autocracy (2 initial poles); 
2T) Society, Culture, Education (1 initial pole); 
3T) Communism, Socialism, Anarchism, Nazism, Fascism (2 initial poles); 
4T) Liberalism, Democracy, Republic (1 initial pole) and 
5T) Capitalism, Money, Economy (1 initial pole).
Thus the natural pole Communism has the largest number of diagram members
even if it is composed of nodes belonging to two different initial poles. 

\def\netwidth{0.7}

\begin{figure}
\centering
\includegraphics[width=\netwidth\textwidth]{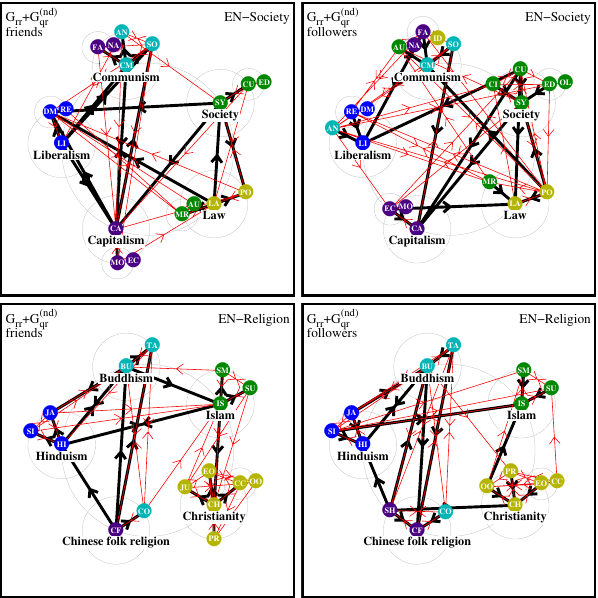}
\caption{\label{fig8} 
Effective friend (left panels) and follower (right panel) network diagrams 
generated 
from the society sub-block of $\Grr+\Gqrnd$ (top panels; 
using the matrix elements $\Grr(i,j)+\Gqrnd(i,j)$ with $i,j\le 23$) 
and from the religion sub-block of $\Grr+\Gqrnd$ (bottom panels; 
using the matrix elements $\Grr(i,j)+\Gqrnd(i,j)$ with $i,j\ge 24$), 
both corresponding to the Wikipedia edition EN. For details about 
the construction method of these diagrams see the text at the beginning 
of Subsection \ref{subsec4.3}. 
The five label colors olive, green, cyan, blue and indigo correspond 
to the pole index 1, 2, 3, 4 and 5 respectively. 
The two character node labels (or codes) and the pole index attribution 
to the nodes are defined in Table~\ref{table2}. 
}
\end{figure}

The diagram of followers has the natural poles 
1T) Law, Politics, Monarchy (2 initial poles); 
2T) Society, Culture, Education, Civilization, Oligarchy (1 initial pol); 
3T) Communism, Socialism, Fascism, Nazism, Autocracy,  Idealism (5 initial poles); 
4T) Liberalism, Democracy, Republic, Anarchism (2 initial pol); 
5T) Capitalism, Money, Economy (1 initial pol). 
Thus again the strongest natural pole is formed around Communism.

For the diagram of religion friends we have from Figure~\ref{fig8} 
the natural pole members: 
1T) Christianity, Catholic Church, Eastern Orthodox Church,
Judaism, Protestantism, Oriental Orthodox Churches (1 initial pol); 
2T) Islam, Sunni Islam, Shia Islam (1 initial pol); 
3T) Buddhism, Taoism (1 initial pol); 
4T)  Hinduism, Jainism, Sikhism (1 initial pol); 
5T) Chinese folk religion, Confucianism (2 initial poles). 
The strongest pole is Christianity, however, it is somehow
isolated having strong links only from Islam
while the poles of Islam, Buddhism, Hinduism, Chinese folk religion
have more active interconnections.

For the diagram of religion followers we find:
1T) Christianity,  Eastern Orthodox Church, Protestantism, Oriental Orthodox Churches, Catholic Church (1 initial pole); 
2T) Islam, Sunni Islam, Shia Islam (1 initial pole); 
3T) Buddhism, Taoism (1 initial pole); 
4T)  Hinduism, Jainism, Sikhism (1 initial pole); 
5T) Chinese folk religion, Confucianism,  Shinto (2 initial poles). 
Here the strongest pole is again Christianity,
and now it is less isolated with connections to Islam and 
Chinese folk religion. At the same time we see here more intense 
links between religions from Asia (3T, 4T, 5T)
forming a strongly interconnected religion group.

\begin{figure}
\centering
\includegraphics[width=\netwidth\textwidth]{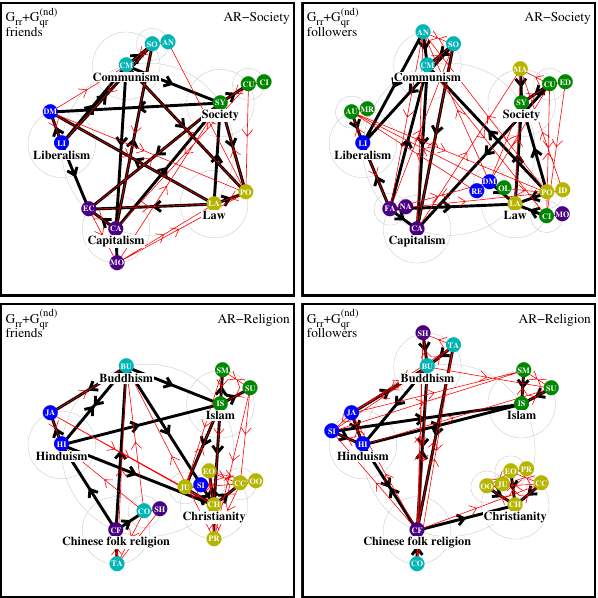}
\caption{\label{fig9} 
As Figure~\ref{fig8} for the Wikipedia edition AR. 
}
\end{figure}

The diagram of friends for society concepts of AR, whose 
network diagrams are shown Figure~\ref{fig9}, 
has a reduced number of nodes as compared to EN in Figure~\ref{fig8},
but the poles are more interconnected by strong links.
Interestingly, the Communism pole does not include Fascism, Nazism
in contrast to the EN edition. For the diagram of society followers
the natural pole with largest number of nodes is Law with 7 members 
and 4 initial poles, 
The Communism pole includes only Socialism and Anarchism
in contrast to EN where the (natural) pole of Capitalism also 
includes Fascism and Nazism.

For the AR diagram of friends for religions in Figure~\ref{fig9}, 
we see that the Christianity pole contains a larger number of members 
as in the EN case but there are more links between poles and the 
Christianity pole is not as isolated as it is for EN. 
However, for the diagram of followers the 
Christianity pole remains more isolated compared to the EN, 
also there are no strong
black links between Christianity and Islam.

\begin{figure}
\centering
\includegraphics[width=\netwidth\textwidth]{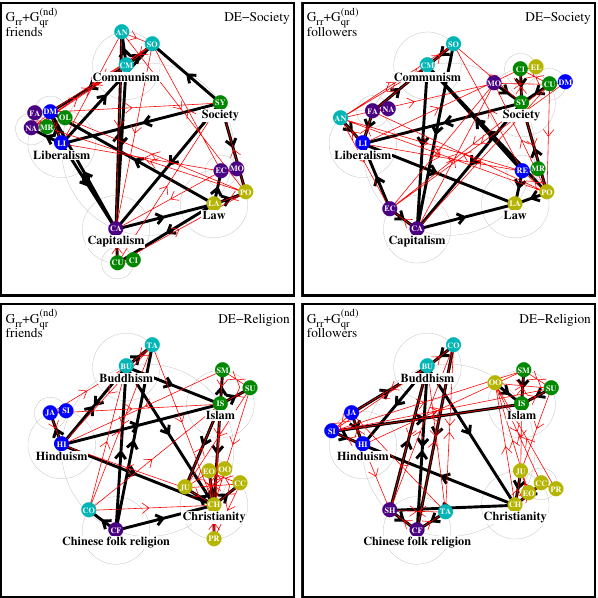}
\caption{\label{fig10} 
As Figure~\ref{fig8} for the Wikipedia edition DE. 
}
\end{figure}

For the DE edition the diagrams are presented in Figure~\ref{fig10}.
For the diagram of friends for society concepts
the strongest natural pole is Liberalism with Democracy, Fascism, Nazism,
Monarchy and Oligarchy (3 initial poles)
while for EN Fascism and Nazism are included in the Communism pole;
also for DE the poles are more densely interconnected as compared to EN.
For the diagram of followers, we again see a significant difference with EN,
thus Fascism and Nazism are included in Liberalism pole while they 
are in the Communism pole for EN. 

For the DE religion diagrams we have denser interconnections between 
the 5 poles as compared to EN. In the case of followers there are
no strong links between Islam and Christianity but there are many 
(level-2) red links between their respective pole members. 

\begin{figure}
\centering
\includegraphics[width=\netwidth\textwidth]{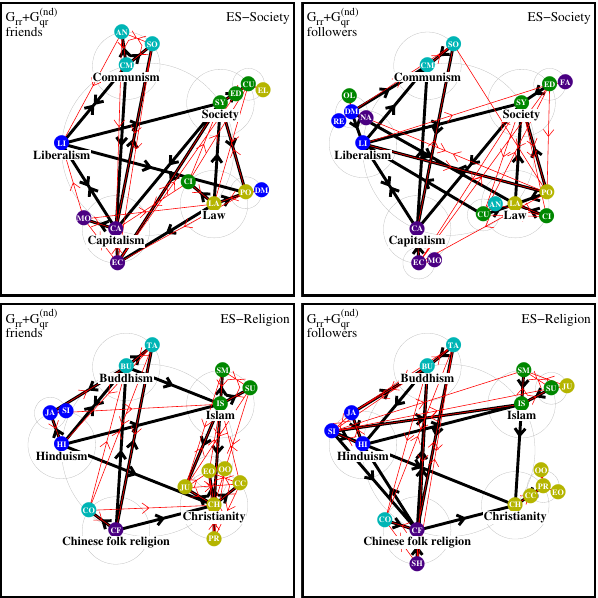}
\caption{\label{fig11} 
As Figure~\ref{fig8} for the Wikipedia edition ES. 
}
\end{figure}

For the ES edition the network diagrams are given in Figure~\ref{fig11}.
Its society friend diagram is similar to EN but there are less
nodes in the Liberalism pole (i.~e. the direct friends of Liberalism 
are the other four level-1 top nodes), also Fascism and Nazism are absent.
For the case of followers there are less pole members
for Society, Communism but more for Law and Liberalism;
Nazism is attributed to the Liberalism pole, Nazism is absent
which is different from the EN case. 

For the religion diagrams of ES in Figure~\ref{fig11}
the case of friends is similar to those of the EN edition
but there are less links between the Islam and Christianity poles.

\begin{figure}
\centering
\includegraphics[width=\netwidth\textwidth]{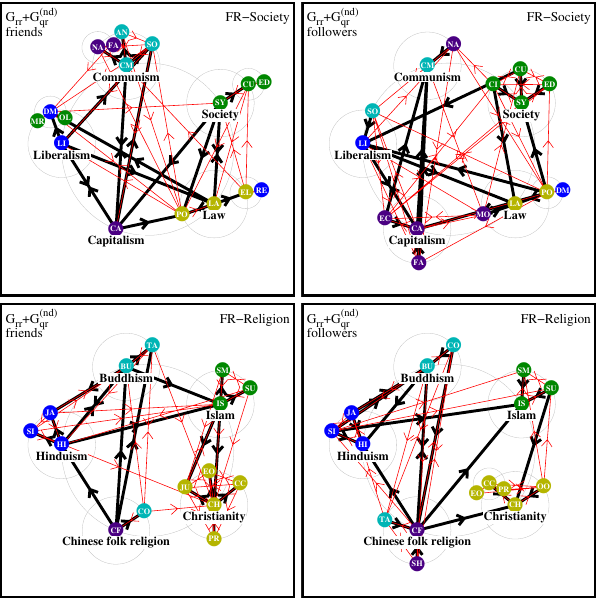}
\caption{\label{fig12} 
As Figure~\ref{fig8} for the Wikipedia edition FR. 
}
\end{figure}

For the FR edition the network diagrams are shown in Figure~\ref{fig12}.
Here the society friend diagram is similar to the case of EN but with less 
links between the Society and Liberalism poles; 
as for EN the nodes Nazism and Fascism belong to the Communism natural pole.
For the case of followers the Communism pole has only one node Nazism 
(from another initial pole) 
while for EN this pole contains 6 members including Fascism and Nazism.
The religion diagrams of FR are quite similar with those of EN. 

\begin{figure}
\centering
\includegraphics[width=\netwidth\textwidth]{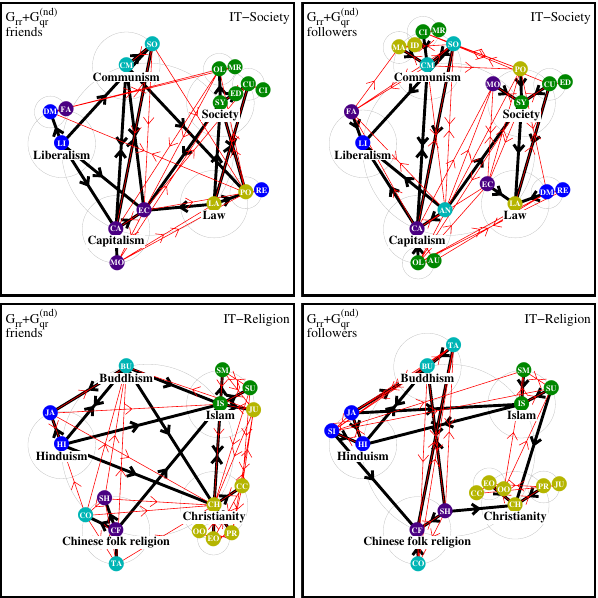}
\caption{\label{fig13} 
As Figure~\ref{fig8} for the Wikipedia edition IT. 
}
\end{figure}

The network diagrams for the IT edition are presented in Figure~\ref{fig13}. 
For the society friend diagram the largest pole is Society including
Culture, Education, Civilization, Oligarchy and Monarchy (all nodes from 
the same initial pole); 
the Liberalism pole includes Democracy and Fascism while Communism 
has only Socialism that makes the last two poles rather different from 
the EN edition;
the Capitalism pole is the same as for EN case; the 
Law pole contains only Politics and Republic.
For the society follower network of IT the highest number of members 
is in the pole of Communism including Socialism, Materialism, Idealism, 
Civilization and Monarchy (3 initial poles).
In both society friend and follower diagrams, thoe node Fascism is 
attributed to Liberalism and while Nazism is absent which constitutes 
a drastic difference with the EN case. 

In the religion friend diagram of IT the node 
Christianity has the highest number of nodes
and it is strongly linked with Islam, Buddhism and Hinduism
in contrast to EN where this pole is more isolated.
The diagram of followers is similar to those of the EN edition
and Christianity remains the strongest pole.

\begin{figure}
\centering
\includegraphics[width=\netwidth\textwidth]{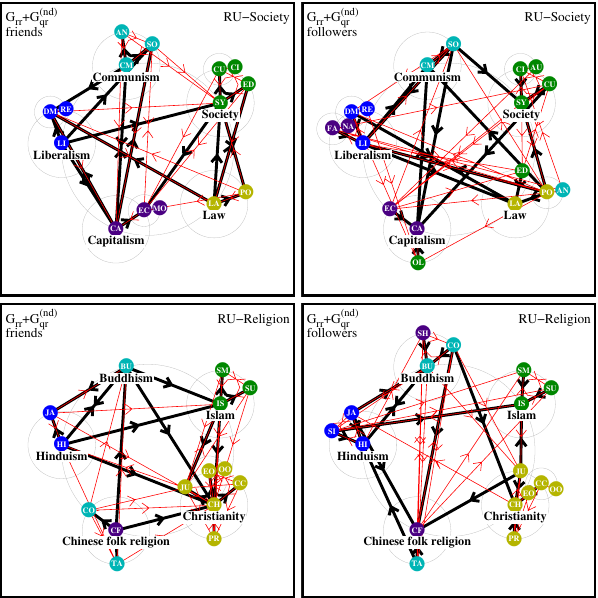}
\caption{\label{fig14} 
As Figure~\ref{fig8} for the Wikipedia edition RU. 
}
\end{figure}

In Figure~\ref{fig14} we show the network
diagrams for the RU edition. Here the society friend diagram is similar 
to EN but without Fascism, Nazism 
in the diagram, also poles Law and Society
have a bit less of included nodes. In the diagram of followers
Fascism and Nazism are included in the Liberalism pole
in contrast to the EN edition where these 2 nodes
are included in Communism pole.

For the religion friend diagram 
Christianity has the largest number of nodes
including Judaism linked also from Islam and 
this pole is less isolates as in the EN edition. 
For the diagram of religion followers
Christianity is still the largest pole with 6 nodes including Judaism
pointing to Islam but in other aspects this diagram is similar to the 
EN edition. 

\begin{figure}
\centering
\includegraphics[width=\netwidth\textwidth]{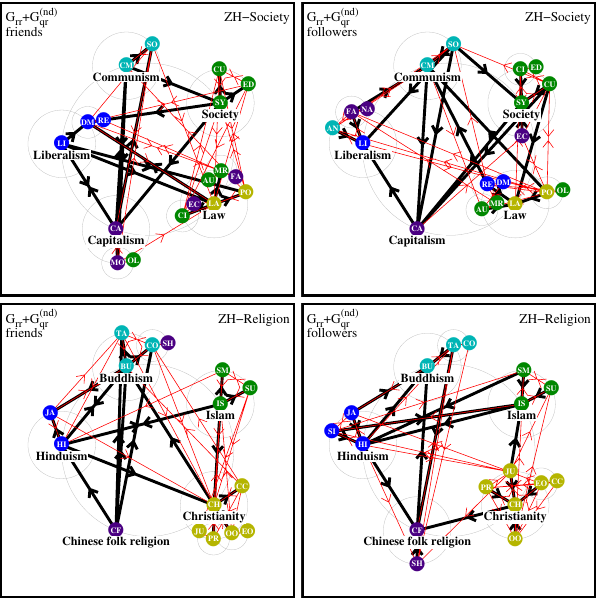}
\caption{\label{fig15} 
As Figure~\ref{fig8} for the Wikipedia edition ZH. 
}
\end{figure}

Finally, for the edition ZH the diagrams are presented in Figure~\ref{fig15}.
In the society friend diagram 
the strongest pole is Law including Politics, Civilization,
Autocracy, Monarchy, Economy and surprisingly Fascism.
We note that for ZH the node Law has the unusual local Rank value $K=3$ 
for a society node which are normally well behind the religion nodes 
in PageRank order. 
The Communism pole includes only Socialism being well linked to the 
Society pole, in contrast to the EN case; Liberalism and Capitalism
poles are similar to EN. 
In the society followers diagram the strongest pole is again 
Law with 6 nodes and 3 initial poles; 
Fascism and Nazism are included in the Liberalism pole.

For the religion friend diagram of ZH the strongest pole is Christianity with
6 nodes being well connected to other poles, in contrast to the EN case;
at the same time the interlinks between Asian religion poles
Buddhism, Hinduism, Chinese folk religion are denser as compared to EN. 
In the religion follower diagram the strongest pole is also Christianity 
with 6 nodes, the diagram structure is similar to EN 
with a larger number of links between Islam and Hinduism poles.

\subsection{Proximity and differences of cultures}

Let us summarize the most important differences and similarities 
between the 8 cultures represented by the 8 language Wikipedia editions 
obtained in the last subsection by analyzing the different 
network diagrams. 

First for the society diagrams the English and French cultures attribute 
the two nodes Fascism and Nazism to the Communism pole while 
they are attributed to the Capitalism pole by the Arabic culture 
and to the Liberalism pole by the German, Spanish (partially), 
Italian (partially), Russian (for followers) and Chinese (for followers) 
cultures. 

Concerning the religion diagrams, the Christianity pole seems to be 
rather isolated in the English culture 
with other links only from the Islam pole (in the friend diagram). 
On the other hand, for the other cultures the Christianity 
pole is well connected not only with the Islam pole
but also with the other poles of Hinduism, Buddhism and
Chinese folk religion. For a majority of cultures
the three poles of the above Asian religions
have a higher density of links between them
as compared to the Islam and Christianity poles.

\begin{figure}
\centering
\includegraphics[width=\netwidth\textwidth]{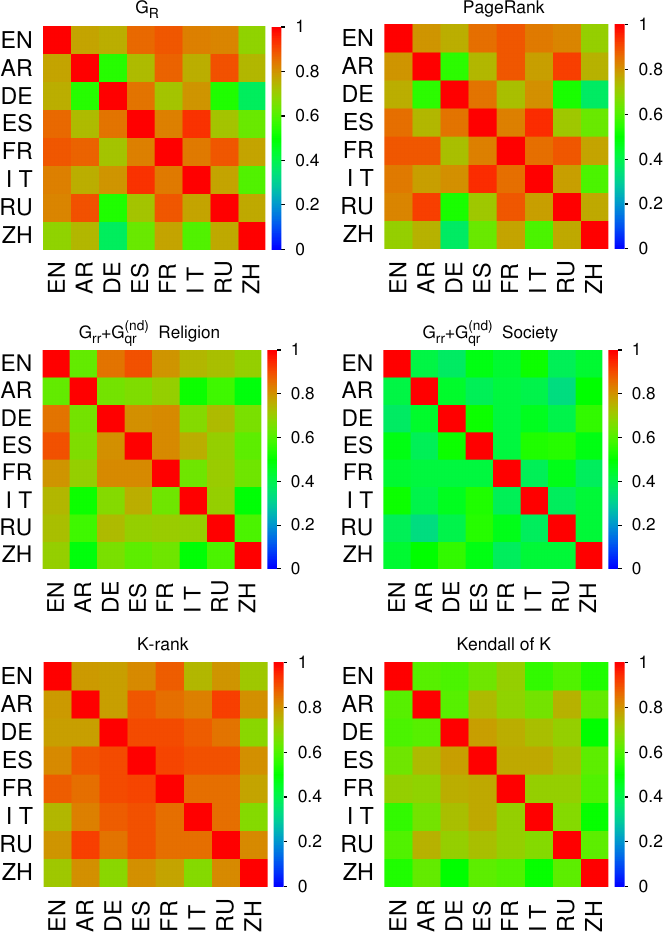}
\caption{\label{fig16}
Color density plots of correlator between the 8 
Wikipedia editions of Table~\ref{table1} for different quantities.
Both top, both center and bottom left panels correspond to the 
Pearson correlator (\ref{defcorr1}) of the five quantities mentioned 
in Subsection \ref{subsec3.4} (and also indicated in the panel 
titles) and the bottom right panel corresponds to the Kendall rank correlator 
(\ref{defcorrKendall}) for the local PageRank index $K$. 
The values of the color bar indicate the correlator value. 
Since no negative correlator values appear only 
a color bar for positive values in the interval $[0,1]$ is shown in all 
cases. The minimal correlator values for the 6 panels (left to right and 
top to bottom) are 0.381, 0.375, 0.486, 0.332, 0.674, 0.5 and 
the maximal off-diagonal correlator values are 
0.933, 0.939, 0.892, 0.565, 0.918, 0.782.
}
\end{figure}

To determine quantitatively the proximity of the 8 cultures, 
we compute the correlators for certain key quantities 
shown in Figure~\ref{fig16}. The 6 panels 
of Figure~\ref{fig16} provide $8\times 8$-matrix density plots for 
different inter edition correlators with 5 panels for the 
Pearson correlator of 5 quantities being the matrix $\GR$, the (group local) 
PageRank vector $\PR$, the religion and society sub-blocks for 
$\Grr+\Gqrnd$ and the local PageRank index $K$ and one additional 
panel for the Kendall correlator of $K$. 
The precise definitions of these correlator quantities with some 
additional technical details are given in Subsection~\ref{subsec3.4} 
and we note that for such correlator 
quantities the minimal mathematical possible value is $-1$, for the case 
of two data sets with strong anti-correlations, while values close 
to $0$ indicate weak or absent correlations and values close to $+1$ 
correspond to strong correlations. 

First, we observe that generally all 8 Wikipedia edition seem to be 
rather well correlated with a big majority of correlator values being 
above 0.5 and  only a few values close to 0.33. 
The two correlators associated to $\GR$ and $\PR$ (top row 
of Figure~\ref{fig16}) are very close which 
is plausible due to the typical strong numerical weight of $\Gpr$ in $\GR$ 
and the fact that the columns of $\Gpr$ are close to $\PR$. Here 
the correlations of DE between AR, RU and ZH seem minimal (still 
with values $\sim 0.5$) and also ZH seems to be less correlated to 
the other editions (with some fluctuations). The other inter edition 
correlations are typically quite strong $\sim 0.8$ with 
the largest values $0.93$-$0.94$ for the correlation between ES and IT. 

More specifically, for the $\GR$-correlator and EN the closest other 
editions are FR (0.882) and ES (0.858); for AR the largest values 
are with RU (0.896) and FR (0.867) which appears to be plausible due to the, 
at least partial, importance of Islam in these three cultures. For 
DE the two closest cultures are ES (0.845) and IT (0.803); for ES they 
are IT (0.933) and EN 0.858); for FR they are RU (0.885)  EN (0.882); 
for IT they are ES (0.933) and FR (0.842); for RU they are 
AR (0.896) and FR(0.885) and finally for ZH they are FR (0.774) and 
RU (0.767). For the very similar $\PR$-correlator this list 
of closest two editions is identical with only 
slightly different correlator values.

Concerning the religion block of $\Grr+\Gqrnd$ (center left panel 
of Figure~\ref{fig16}), we see that AR and ZH have globally the 
weakest correlations to the other cultures, with values $\sim 0.5$, 
which seems natural due to the importance of their 
specific religions. On the other hand here, we have a block of 
four strongly correlated editions EN, DE, ES, FR between them, 
with values $\ge 0.8$, while IT and RU have typical ``intermediate'' 
correlations $\sim 0.7$. 

More explicitly, for this case the closest cultures of EN are 
ES (0.892) and DE (0.844); for AR they are FR (0.699) and ES (0.662) 
with relatively low values; for DE they are EN (0.844), FR (0.814) 
and ES (0.806); for ES they are  EN (0.892), FR (0.814) and DE (0.806); 
for FR they are DE (0.814), ES (0.814) and EN (0.801); 
for IT they are ES (0.76) and EN (0.746); for RU they are 
DE (0.737) and EN (0.73) and finally the strongest correlator of ZH 
is to EN (0.694). 

The society block of $\Grr+\Gqrnd$ (center right panel 
of Figure~\ref{fig16}) shows the ``weakest'' general 
correlations of all correlator quantities with (off-diagonal) values 
being typically $\sim 0.5$ with the strongest off-diagonal element 
being 0.565 between DE and ZH which is due to two strong matrix elements 
in $\Grr+\Gqrnd$ due to the links from Oligarchy to Democracy and Monarchy 
for both editions. Here 
the AR-RU correlator represents the minimal correlator value $0.332$ for all 
editions and all correlator quantities. 

The Pearson and Kendall correlator of the PageRank index $K$ 
(bottom row of Figure~\ref{fig16}) appear to have roughly a similar relative 
structure as the Pearson 
correlators of $\GR$ and $\PR$ (top row of Figure~\ref{fig16}). However, 
here the overall values are significantly stronger (lower) for 
the Pearson (Kendall) $K$-correlator in comparison to the top row values. 
For the Kendall correlator of $K$ and the two editions EN and ZN there is 
an additional suppression of the correlator values to other editions 
which are mostly close to $\sim 0.5$. Furthermore, for both $K$-correlators, 
we have a block of 5 editions DE, ES, FR, IT and RU of relatively strong 
correlations between them and AR has somewhat intermediate correlations 
to this block while EN and ZH seem to be a bit more separated from this 
block (but still with significant correlator values). 

\section{Discussion and conclusion}
\label{sec5}

In this work, we presented the Google matrix analysis of Wikipedia networks 
constructed from 8 language editions
(EN, AR, DE, ES, FR, IT, RU, ZH) collected at 1 October 2024 and with 
key properties given in Table~\ref{table1}. 
Specifically, we analyzed the relations and interactions between
40 article entries about 23 society concepts and 17 religions or 
branches of religions (see Tables~\ref{table2} and \ref{table3}). 
Using the PageRank and CheiRank vectors it is possible 
to establish a ranking of these 40 articles based 
on either their importance or their communicativeness respectively. 
We find that globally in this group the articles related to religion 
are located at higher PageRank positions, implying higher importance, 
than the society related articles including 
Law, Society, Liberalism, Capitalism, Communism etc. Exceptions 
are the articles of Nazism with 2nd PageRank position in the German 
edition and Law with 3rd PageRank position in the Chinese edition. 

Using the established REGOMAX algorithm \cite{politwiki}, 
we computed for each edition the reduced Google matrix $\GR$ and its 
components which describe the direct and indirect transitions
between all 40 entries (nodes), the latter taking into account all
indirect pathways using nodes outside the group of 40 articles 
via the huge global matrix of the whole Wikipedia network. 
We find that the two diagonal blocks for society and religion nodes 
of the matrix $\Grr + \Gqrnd$, representing direct and 
``interesting'' indirect links, nearly decouple with 
significantly smaller transitions between these two blocks. 

Therefore the interactions between society concepts and religions
are relatively weak for all 8 editions
even if the historical role
of religions on society development is well known.
We conjecture that this is partially related to 
the well known Bible statement 
{\it Render unto Caesar the things that are Caesar's, and unto God the things that are God's}
(Bible Matthew 22:21 \cite{bible})
but there may be also other reasons.

We also extracted effective network friend and follower diagrams 
from the two diagonal blocks of $\Grr + \Gqrnd$ 
providing a compact description of relations inside either the sector of
society concepts or inside the sector of religions. 
For example, depending on the edition, 
the concepts of Fascism and Nazism are attributed 
to different influence poles such as Communism (EN, FR),
Liberalism (DE, IT, RU, ZH) or Capitalism (AR).
For the sector of religions we note that for some editions the 
Christianity pole is rather isolated from other religion  poles (e.g. EN) 
while for most other editions it is well connected to other
poles of Buddhism, Hinduism, Islam, Chinese folk religion.
For a majority of editions the links between Asian religions
represented by the poles of Buddhism, Hinduism and Chinese folk religion
are stronger than the links of the two Christianity and Islam poles. 

Finally, we also provided a quantitative analysis of inter edition 
correlators for various key quantities (reduced Google matrix components or 
blocks, PageRank vector or Index etc.) which allows to determine 
the proximity or distance of different cultures, represented by 
the Wikipedia language editions, with respect to their views 
on the 40 selected Wikipedia articles. 
For example, for $\GR$ the Arabic (Chinese) culture has 
the strongest correlations with the Russian (French) culture and 
the German edition is closest to the Spanish and Italian editions. 
If we consider the religion diagonal block of $\Grr + \Gqrnd$, we 
have the strongest culture proximity between EN ans ES. Generally speaking 
the overall inter edition correlations are rather large, with most values 
above $0.5$ often close to 0.8-0.9 and only a few minimal values close 
to 0.33.

In conclusion, we presented a mathematical network analysis of relations 
and interactions of 23 society concepts
and 17 religions for 8 Wikipedia editions 
allowing to extract nontrivial features of these relations.
We note that the described approach can be applied to 
any selected subset (topic) of modest size of Wikipedia articles. 

\section*{Appendix}
\setcounter{equation}{0}
\renewcommand{\theequation}{A\arabic{equation}}
\setcounter{figure}{0}
\setcounter{section}{0}
\renewcommand\thefigure{A\arabic{figure}}
\renewcommand\thesection{A\arabic{section}}
\renewcommand{\figurename}{Appendix Figure}


\authorcontributions{All authors equally contributed to all stages of this work.
}

\funding{The authors acknowledge support from the grant
 ANR France project
NANOX $N^\circ$ ANR-17-EURE-0009 in the framework of 
the Programme Investissements d'Avenir (project MTDINA).
}




\acknowledgments{We thank L.Ermann for useful discussions.
}

\conflictsofinterest{The authors declare no conflict of interest.
} 



\reftitle{References}

\end{document}

%% file: table1.tex
\begin{tabular}{llrrrr}
\toprule
Language & Edition & $N$ & $N_\ell$ & $N_d$ & $N_\ell/N$ \\
\midrule
English & EN & 6891535 & 185658675 & 18444 & 26.9 \\
Arabic & AR & 1242011 & 16433487 & 20467 & 13.2 \\
German & DE & 2946636 & 79189123 & 20099 & 26.9 \\
Spanish & ES & 1916240 & 41324254 & 4243 & 21.6 \\
French & FR & 2638634 & 76118849 & 3567 & 28.8 \\
Italian & IT & 1884339 & 49495890 & 7622 & 26.3 \\
Russian & RU & 2002167 & 43375388 & 7630 & 21.7 \\
Chinese & ZH & 1444719 & 20682593 & 58250 & 14.3 \\
\bottomrule
\end{tabular}

%% file: table2.tex
\begin{tabular}{rrrllrrl}
\toprule
$K_g$ & $K$ & $K^*$ & Title & Code & $K_{\rm M}$ & $K_{\rm M}^*$ &subgroup\\
\midrule
1 & 7 & 28 & Law & LA& 387 & 101448 & 1(T)\\
2 & 10 & 36 & Education & ED& 531 & 323268 & 2\\
3 & 11 & 15 & Communism & CM& 690 & 26967 & 3(T)\\
4 & 12 & 19 & Democracy & DM& 750 & 41331 & 4\\
5 & 13 & 16 & Liberalism & LI& 755 & 28568 & 4(T)\\
6 & 14 & 21 & Socialism & SO& 837 & 42483 & 3\\
7 & 16 & 20 & Ecology & EL& 988 & 41611 & 1\\
8 & 17 & 3 & Politics & PO& 1072 & 4444 & 1\\
9 & 18 & 9 & Culture & CU& 1190 & 10959 & 2\\
10 & 19 & 22 & Nazism & NA& 1234 & 49141 & 5\\
11 & 20 & 30 & Capitalism & CA& 1282 & 117558 & 5(T)\\
12 & 22 & 34 & Republic & RE& 1652 & 283031 & 4\\
13 & 25 & 11 & Monarchy & MR& 1903 & 15001 & 2\\
14 & 27 & 27 & Fascism & FA& 2089 & 89729 & 5\\
15 & 28 & 4 & Society & SY& 2448 & 6032 & 2(T)\\
16 & 29 & 33 & Economy & EC& 2560 & 261158 & 5\\
17 & 31 & 5 & Anarchism & AN& 3663 & 8827 & 3\\
18 & 33 & 39 & Money & MO& 4624 & 425841 & 5\\
19 & 34 & 40 & Oligarchy & OL& 5759 & 1200109 & 2\\
20 & 35 & 8 & Civilization & CI& 6035 & 9853 & 2\\
21 & 36 & 37 & Autocracy & AU& 6074 & 324067 & 2\\
22 & 38 & 35 & Materialism & MA& 7709 & 316619 & 1\\
23 & 40 & 31 & Idealism & ID& 11579 & 130647 & 1\\
\midrule
24 & 1 & 6 & Catholic Church & CC& 60 & 8870 & 1\\
25 & 2 & 2 & Christianity & CH& 85 & 1307 & 1(T)\\
26 & 3 & 7 & Islam & IS& 96 & 9758 & 2(T)\\
27 & 4 & 17 & Buddhism & BU& 188 & 28678 & 3(T)\\
28 & 5 & 10 & Hinduism & HI& 255 & 11219 & 4(T)\\
29 & 6 & 18 & Eastern Orthodox Church & EO& 353 & 31721 & 1\\
30 & 8 & 12 & Judaism & JU& 401 & 16492 & 1\\
31 & 9 & 13 & Protestantism & PR& 405 & 22902 & 1\\
32 & 15 & 24 & Sunni Islam & SU& 890 & 63317 & 2\\
33 & 21 & 1 & Jainism & JA& 1361 & 275 & 4\\
34 & 23 & 25 & Sikhism & SI& 1659 & 64136 & 4\\
35 & 24 & 32 & Confucianism & CO& 1794 & 139315 & 3\\
36 & 26 & 14 & Shia Islam & SM& 1945 & 26758 & 2\\
37 & 30 & 23 & Taoism & TA& 2576 & 50597 & 3\\
38 & 32 & 29 & Shinto & SH& 3999 & 116232 & 5\\
39 & 37 & 26 & Chinese folk religion & CF& 6105 & 65071 & 5(T)\\
40 & 39 & 38 & Oriental Orthodox Churches & OO& 9789 & 331323 & 1\\
\bottomrule
\end{tabular}

%% file: table3.tex
\begin{tabular}{rlllllllll}
\toprule
$K_g$ & Code/& EN& AR& DE& ES& FR& IT& RU& ZH\\
 & Node& $K;K^*$& $K;K^*$& $K;K^*$& $K;K^*$& $K;K^*$& $K;K^*$& $K;K^*$& $K;K^*$\\
\midrule
1 & LA & 7;28 & 5;18 & 17;8 & 5;15 & 5;37 & 5;22 & 15;14 & 3;22 \\
2 & ED & 10;36 & 15;3 & 25;40 & 11;30 & 19;33 & 26;28 & 20;26 & 8;7 \\
3 & CM & 11;15 & 16;22 & 13;23 & 18;10 & 10;2 & 12;13 & 16;9 & 13;1 \\
4 & DM & 12;19 & 10;12 & 6;7 & 6;22 & 11;14 & 11;10 & 13;19 & 12;12 \\
5 & LI & 13;16 & 22;16 & 12;22 & 15;35 & 13;13 & 16;14 & 18;28 & 25;19 \\
6 & SO & 14;21 & 21;28 & 18;17 & 21;9 & 18;4 & 15;23 & 10;15 & 16;16 \\
7 & EL & 16;20 & 27;15 & 20;28 & 22;12 & 17;17 & 22;26 & 21;21 & 28;30 \\
8 & PO & 17;3 & 4;26 & 10;25 & 7;31 & 6;11 & 8;31 & 14;37 & 4;28 \\
9 & CU & 18;9 & 17;17 & 8;15 & 12;17 & 14;21 & 10;19 & 12;32 & 7;34 \\
10 & NA & 19;22 & 25;13 & 2;13 & 23;13 & 12;22 & 21;11 & 28;12 & 32;38 \\
11 & CA & 20;30 & 20;21 & 23;10 & 20;23 & 22;16 & 19;20 & 22;11 & 23;10 \\
12 & RE & 22;34 & 18;35 & 21;38 & 17;27 & 8;31 & 20;30 & 19;36 & 15;23 \\
13 & MR & 25;11 & 23;37 & 14;6 & 19;6 & 21;32 & 17;21 & 17;1 & 17;9 \\
14 & FA & 27;27 & 29;31 & 22;11 & 24;5 & 27;6 & 9;3 & 26;8 & 33;20 \\
15 & SY & 28;4 & 7;24 & 19;26 & 16;34 & 26;39 & 18;40 & 11;35 & 22;32 \\
16 & EC & 29;33 & 12;33 & 7;27 & 3;38 & 20;40 & 4;6 & 7;33 & 14;39 \\
17 & AN & 31;5 & 30;7 & 27;4 & 26;1 & 28;1 & 38;24 & 30;3 & 29;5 \\
18 & MO & 33;39 & 26;40 & 26;14 & 25;36 & 24;23 & 29;37 & 24;20 & 20;27 \\
19 & OL & 34;40 & 37;38 & 31;39 & 33;39 & 30;35 & 31;38 & 36;34 & 36;14 \\
20 & CI & 35;8 & 24;8 & 35;36 & 30;21 & 32;27 & 30;33 & 27;31 & 24;25 \\
21 & AU & 36;37 & 36;36 & 36;37 & 37;40 & 37;38 & 37;39 & 39;40 & 30;29 \\
22 & MA & 38;35 & 38;27 & 33;30 & 35;14 & 35;29 & 34;17 & 33;23 & 39;26 \\
23 & ID & 40;31 & 40;25 & 38;33 & 36;37 & 38;10 & 32;32 & 35;39 & 38;36 \\
\midrule
24 & CC & 1;6 & 3;14 & 1;16 & 1;2 & 3;26 & 1;1 & 4;27 & 10;13 \\
25 & CH & 2;2 & 2;1 & 3;9 & 2;3 & 1;12 & 2;4 & 2;4 & 1;17 \\
26 & IS & 3;7 & 1;2 & 4;1 & 4;4 & 2;3 & 3;5 & 1;7 & 5;11 \\
27 & BU & 4;17 & 11;11 & 11;3 & 9;7 & 7;5 & 25;8 & 5;6 & 2;3 \\
28 & HI & 5;10 & 13;10 & 16;5 & 14;8 & 15;7 & 14;9 & 8;5 & 9;24 \\
29 & EO & 6;18 & 19;19 & 15;20 & 13;16 & 16;20 & 13;15 & 23;16 & 34;15 \\
30 & JU & 8;12 & 9;4 & 5;12 & 8;11 & 9;9 & 6;2 & 6;13 & 18;37 \\
31 & PR & 9;13 & 6;9 & 9;19 & 10;18 & 4;15 & 7;18 & 3;18 & 6;31 \\
32 & SU & 15;24 & 8;5 & 24;2 & 27;32 & 23;36 & 23;34 & 9;30 & 21;18 \\
33 & JA & 21;1 & 34;32 & 34;31 & 34;19 & 34;8 & 36;16 & 34;17 & 27;33 \\
34 & SI & 23;25 & 28;20 & 30;21 & 29;20 & 29;18 & 27;27 & 32;10 & 31;35 \\
35 & CO & 24;32 & 32;34 & 28;34 & 32;29 & 33;34 & 33;29 & 31;29 & 19;4 \\
36 & SM & 26;14 & 14;6 & 32;24 & 31;24 & 25;25 & 24;25 & 25;22 & 26;21 \\
37 & TA & 30;23 & 31;29 & 29;18 & 28;25 & 31;19 & 28;12 & 29;24 & 11;2 \\
38 & SH & 32;29 & 35;30 & 37;29 & 38;28 & 36;24 & 35;7 & 38;2 & 35;8 \\
39 & CF & 37;26 & 39;39 & 40;32 & 40;26 & 39;30 & 40;35 & 40;38 & 40;6 \\
40 & OO & 39;38 & 33;23 & 39;35 & 39;33 & 40;28 & 39;36 & 37;25 & 37;40 \\
\bottomrule
\end{tabular}

%% file: wikisoc.bbl
\begin{thebibliography}{99}

\bibitem{jarvie} Jarvie, J.C. {\it Concepts and society},
           Routledge \& Kegan Paul Ltd., London, UK (1972).
\bibitem{gellner} Gellner, E. {\it Cause and meaning in the social sciences},
            Routledge, London, UK (1973).
\bibitem{casanova} Casanovam J.
        {\it Public religions in the modern world},
         University of Chicago Press, Chicago (1980).
\bibitem{reese} Reese, W.L. 
          {\it Dictionary of philosophy and religion: Eastern and Western thought},
            Humanity Books (1996) 
\bibitem{barrett} Barrett, J.L. 
             {\it Exploring the natural foundations of religion},
              Trends in Cognitive Sciences  {\bf 2000}, {\em 4(1)}, 29.
\bibitem{whitehouse} Whitehous, H.; Martin,  L.H. (Eds.),
        {\it Theorizing religions past: Archaelogy, history, and cognition},
        Altamira Press, Walnut Creek, CA (2004).
\bibitem{atran} Atran, S.;  Norenzayan, A.
           {\it Religion's evolutionary landscape: Counterintuition, 
        commitment, compassion, communion},
            Behavior. Brain. Sciences {\bf 2004}, {\em  27(6)}, 713.
\bibitem{boyer} Boyer, P. {\it Religion explained},
        Random House, London (2008).
\bibitem{hopfe} Hopfe, L.M.; Woodward, N.R.
         {\it Religions of the world},
        Vango Books, N. Y. (2009).
\bibitem{britanica} Encyclopaedia Brittanica 
           \url{http://www.britannica.com/} (Accessed 12 November 2024).                  
\bibitem{wiki1}  Giles, J. {\it Internet encyclopaedias go head to head}, 
          Nature {\bf 2005}, {\em 438}, 900. 
\bibitem{reagle} Reagle Jr., J.M. {\it  Good faith collaboration: 
               The culture of Wikipedia},
                 MIT Press, Cambridge MA (2010).
\bibitem{finn} Nielsen, F.A. {\it Wikipedia research and tools: 
          review and comments}, (2012),
         \url{https://papers.ssrn.com/sol3/papers.cfm?abstract\_id=2129874}
           (Accessed 12 November 2024).
\bibitem{abramowicz} Lewoniewski, W.; Wecel, K.; Abramowicz, W.
        {\it Quality and importance of Wikipedia articles in different languages},
        in G.~Dregvaite, R.~Damasevicius (Eds)
         Information and Software Technologies, ICIST 2016,
         Comm. Computer  Inform. Sci. {\bf 2016}, {\em 637}, 613.
\bibitem[Ball(2023)]{wikiacad3} Ball, C.
      \newblock Defying easy categorization: Wikipedia as primary, secondary and tertiary resource.
      \newblock {\em Insights} {\bf 2023}, {\em  7},~1.
      \newblock {\url{DOI: https://doi.org/10.1629/uksg.604}}.
\bibitem[Arroyo-Machado et al.(2024)]{wikiacad4}
        Arroyo-Machado, W., Diaz-Faes, A.A., Herrera-Viedma, E, Castas, R.
        \newblock From academic to media capital: To what extent does the
         scientific reputation of universities translate into
         Wikipedia attention?
         \newblock {\em J. Assoc. Inf. Sci. Technol} {\bf 2024}, {\em  75},~423.
           \newblock {\url{https://asistdl.onlinelibrary.wiley.com/doi/full/10.1002/asi.24856}}.
\bibitem{brin} Brin, S.; Page, L.
         {\it The anatomy of a large-scale hypertextual Web search engine},
         Computer Networks and ISDN Systems {\bf 1998}, {\em 30}, 107.
\bibitem{meyer} Langville, A.M.; Meyer, C.D. 
        {\it  Google's PageRank and beyond: the science of search engine rankings}, 
        Princeton University Press, Princeton (2006).
\bibitem{cheirank} Chepelianskii, A.D.
         {\it Towards physical laws for software architecture},
           arXiv:1003.5455 [cs.SE] (2010).
\bibitem{wikizzs} Zhirov, A.O.; Zhirov, O.V.; Shepelyansky, D.L.
        {\it Two-dimensional ranking of Wikipedia articles},
        Eur. Phys. J. B {\bf 2010}, {\em 77}, 523.
\bibitem{rmp2015} Ermann, L.; Frahm, K.M.; Shepelyansky, D.L.
        {\it Google matrix analysis of directed networks},
          Rev. Mod. Phys. {\bf 2015}, {\em 87}, 1261.
\bibitem{politwiki} Frahm, K.M.; Jaffr\`es-Runser K.; Shepelyansky, D.L.
         {\it Wikipedia mining of hidden links between political leaders},
          Eur. Phys. J. B {\bf 2016}, {\em 89}, 269.
\bibitem{eomwiki24} Eom, Y.-H.; Aragon, P.; Laniado, D.; 
         Kaltenbrunner, A.; Vigna, S.; Shepelyansky, D.L.
        {\it Interactions of cultures and top people of Wikipedia 
         from ranking of 24 language editions},
        PLoS ONE {\bf 2015}, {\bf 10(3)}, e0114825.
\bibitem{wrwu2017} Coquide, C.; Lages, J.; Shepelyansky, D.L. 
         {\it World influence and interactions of universities from Wikipedia networks}, 
         Eur. Phys. J. B {\bf 2019}, {\em 92}, 3.
\bibitem{wikibank}  Demidov, D.; Frahm, K.M,; Shepelyansky, D.L. 
        {\it What is the central bank of Wikipedia?}, 
        Physica A {\bf 2020}, {\em 542}, 123199.
\bibitem{wtn3} Coquide, C.; Ermann, L.; Lages, J.; Shepelyansky, D.L. 
         {\it Influence of petroleum and gas trade on EU economies from 
          the reduced Google matrix analysis of UN COMTRADE data}, 
         Eur. Phys. J. B {\bf 2019}, {\em 92}, 171       .
\bibitem{grmetacore} Kotelnikova, E.; Frahm, K.M.; Shepelyansky, D.L.; Kunduzova. O.  
         {\it Fibrosis protein-protein interactions from Google matrix analysis of MetaCore network},
         Int. J. Mol. Sci. {\bf 2022}, {\em 23}, 67.

\bibitem{ourwebpage} {\it Wikiconcepts.} Available online: 
\url{https://www.quantware.ups-tlse.fr/QWLIB/wikiconcepts/index.html/}
         (Accessed on 3 December 2024).

\bibitem{utf8proc} {\it Utf8proc.} Available online: 
\url{https://juliastrings.github.io/utf8proc}
         (Accessed on 19 October 2024).                  


\bibitem{aragon} P.Aragon, D.Laniado, A.Kaltenbrunner, and Y.Volkovich, 
         {\it  Biographical social networks on Wikipedia: a cross-cultural study of links that made history},
          In: Proceedings of the Eighth Annual International Symposium on Wikis and Open Collaboration, 
          WikiSym ’12, Association for Computing Machinery, New York, NY, USA (2012);
           https://doi.org/10.1145/2462932.2462958 .

\bibitem{schur} {\it The Schur complement and its applications},
                  Ed. Fushen Zhang, Springer, Berlin (2005).
\bibitem{meyer2} Meyer C.D.,
                 {\it Stochastic complementation, uncoupling Markov chains,
                  and the theory of nearly reducible systems},
                  SIAM Review {\bf 1989}, {\em 31(2)}, 240. 

           
\bibitem{wikicorr1} {\it Pearson correlation coefficient.} Available online: 
\url{https://en.wikipedia.org/wiki/Pearson_correlation_coefficient} 
         (Accessed on 7 November 2024).                  

\bibitem{wikicorr2} {\it Kendall rank correlation coefficient.} Available online: 
\url{https://en.wikipedia.org/wiki/Kendall_rank_correlation_coefficient}
         (Accessed on 7 November 2024).                  


\bibitem{bible} {Wikipedia contributors} (2024),
\newblock Render unto Caesar --- {Wikipedia}{,} The Free Encyclopedia.
\newblock
  \url{https://en.wikipedia.org/wiki/Render_unto_Caesar}, 2024.
  \newblock [Online; accessed 18-November-2024].
           

\end{thebibliography}
